\documentclass[11pt,letterpaper]{article}

\usepackage[utf8]{inputenc}
\usepackage[T1]{fontenc}
\usepackage{libertine}        
\usepackage[scaled=0.95]{inconsolata}  

\usepackage{amsmath}
\usepackage[libertine,vvarbb]{newtxmath}  
\usepackage{nicefrac}

\usepackage{geometry}
\usepackage{microtype}        
\usepackage[compact]{titlesec}  

\usepackage{graphicx}
\usepackage{booktabs}         
\usepackage{multirow}
\usepackage{array}

\usepackage{enumitem}
\usepackage{xspace}
\usepackage{pifont}

\usepackage{xcolor}
\usepackage{url}
\usepackage{hyperref}         
\usepackage[capitalize,noabbrev]{cleveref}  

\usepackage[font=small,labelfont=bf]{caption}  
\usepackage{subcaption}
\usepackage{listings}

\usepackage[numbers,sort&compress]{natbib}

\definecolor{TolBlue}{RGB}{68,119,170}       
\definecolor{TolCyan}{RGB}{102,204,238}      
\definecolor{TolGreen}{RGB}{34,136,51}       
\definecolor{TolYellow}{RGB}{204,187,68}     
\definecolor{TolRed}{RGB}{238,102,119}       
\definecolor{TolPurple}{RGB}{170,51,119}     
\definecolor{TolGrey}{RGB}{187,187,187}      

\definecolor{TableHeader}{RGB}{68,119,170}   
\definecolor{TableAlt}{RGB}{245,247,250}     

\hypersetup{
    colorlinks=true,
    linkcolor=TolBlue,      
    citecolor=TolGreen,     
    urlcolor=TolRed,        
    pdfauthor={Michael J. Bommarito II},
    pdftitle={Binary-30K: A Cross-Platform Dataset for Deep Learning in Binary Analysis and Malware Detection},
    pdfsubject={Binary Analysis, Machine Learning, Malware Detection},
    pdfkeywords={binary analysis, malware detection, machine learning, transformers, dataset}
}

\newcommand{\datasetname}{Binary-30K\xspace}
\newcommand{\datasetsplitsurl}{https://huggingface.co/datasets/mjbommar/binary-30k}
\newcommand{\datasettokenizedurl}{https://huggingface.co/datasets/mjbommar/binary-30k-tokenized}
\newcommand{\tokenizerurl}{https://huggingface.co/mjbommar/binary-tokenizer-001-64k}
\newcommand{\binarybpepaperurl}{https://arxiv.org/abs/2511.17573}
\newcommand{\datasetrepourl}{https://github.com/mjbommar/binary-dataset-paper}
\newcommand{\numrecords}{29,793}
\newcommand{\numunique}{29,793}
\newcommand{\nummalware}{8,023}
\newcommand{\malwareratio}{26.93\%}

\newcommand{\numwindows}{17,239}
\newcommand{\numlinux}{8,452}
\newcommand{\nummacos}{568}
\newcommand{\numandroid}{164}
\newcommand{\numother}{3,370}
\newcommand{\numunknownarch}{1,582}
\newcommand{\numnotapplicable}{2,049}
\newcommand{\numexotic}{1,498}
\newcommand{\windowspct}{57.86\%}
\newcommand{\linuxpct}{28.37\%}
\newcommand{\otherpct}{11.31\%}
\newcommand{\macospct}{1.91\%}
\newcommand{\androidpct}{0.55\%}

\newcommand{\sorelscale}{\mbox{$\approx$20 million}}
\newcommand{\soreldisarmed}{\mbox{$\approx$10 million}}
\newcommand{\emberscale}{1.1 million}

\newcommand{\rawsize}{13.18~GiB}

\newcommand{\compressedsize}{approximately 11~GiB}

\graphicspath{{images/}}

\setlist{
    itemsep=0.25em,
    parsep=0.25em,
    topsep=0.5em
}

\titleformat{\section}
    {\Large\bfseries}                    
    {\thesection}                        
    {1em}                                
    {}                                   
    [\vspace{-0.5em}]                   

\titleformat{\subsection}
    {\large\bfseries}                    
    {\thesubsection}                     
    {1em}                                
    {}                                   
    [\vspace{-0.3em}]                   

\titleformat{\subsubsection}
    {\normalsize\bfseries}               
    {\thesubsubsection}                  
    {1em}                                
    {}                                   
    []                                   

\titlespacing*{\section}{0pt}{1.5ex plus 1ex minus .2ex}{1ex plus .2ex}
\titlespacing*{\subsection}{0pt}{1.2ex plus .8ex minus .2ex}{0.8ex plus .2ex}
\titlespacing*{\subsubsection}{0pt}{1ex plus .5ex minus .2ex}{0.5ex plus .2ex}

\title{Binary-30K: \\A Heterogeneous Dataset for Deep Learning \\in Binary Analysis and Malware Detection}

\author{
    Michael J. Bommarito II\thanks{Portions of this work were prepared with assistance from large language models. The author is solely responsible for all content, including any errors or omissions.} \\
    \texttt{michael.bommarito@gmail.com}
}

\date{}

\begin{document}

\maketitle

\begin{abstract}
	Deep learning research for binary analysis faces a critical infrastructure gap. Today, existing datasets target single platforms, require specialized tooling, or provide only hand-engineered features incompatible with modern neural architectures; no single dataset supports accessible research and pedagogy on realistic use cases.  To solve this, we introduce \datasetname{}, the first heterogeneous binary dataset designed for sequence-based models like transformers.  Critically, \datasetname{} covers Windows, Linux, macOS, and Android across 15+ CPU architectures. With \numrecords{} binaries and approximately \malwareratio{} malware representation, \datasetname{} enables research on platform-invariant detection, cross-target transfer learning, and long-context binary understanding. The dataset provides pre-computed byte-level BPE tokenization~\cite{bommarito2025binarybpe} alongside comprehensive structural metadata, supporting both sequence modeling and structure-aware approaches. Platform-first stratified sampling ensures representative coverage across operating systems and architectures, while distribution via Hugging Face with official train/validation/test splits enables reproducible benchmarking. The dataset is publicly available at \href{\datasetsplitsurl}{\texttt{mjbommar/binary-30k}}, providing an accessible resource for researchers, practitioners, and students alike.
\end{abstract}


\section{Introduction}
\label{sec:introduction}

Deep learning methods for binary analysis require datasets that reflect the heterogeneous nature of real-world computing environments: diverse platforms (Windows, Linux, macOS, Android), architectures (x86, ARM, MIPS, RISC-V, and more), and temporal ranges spanning compiler evolution and threat landscapes. \datasetname is a heterogeneous binary dataset purpose-built for deep learning research in malware detection and binary analysis. It provides \numrecords{} pre-tokenized binaries with byte-level BPE encoding using the Binary BPE tokenizer family~\cite{bommarito2025binarybpe}, comprehensive metadata, and distribution via Hugging Face in transformer-ready Arrow/Parquet formats with official stratified train/validation/test splits. The dataset combines this heterogeneous coverage (Windows, Linux, macOS, Android platforms; 15+ CPU architectures; 2012–2025 temporal span) with a realistic \malwareratio{} malware share and pre-computed tokenization specifically designed to enable cross-platform transfer learning, architecture-invariant detection research, and long-context transformer modeling without custom preprocessing infrastructure.

The design emphasizes accessibility for both research and education. The dataset spans 13 years (2012–2025): benign binaries cover OS releases from Windows 8 (October 2012) through Ubuntu 24.04 (April 2024), capturing compiler technology evolution, security hardening adoption (ASLR, DEP, CFG), and build toolchain modernization; malware samples span 2017–2025, combining historical SOREL-20M Windows threats (2017–2019) with contemporary cross-platform Malware Bazaar samples (2020–2025). This temporal depth supports longitudinal analysis of malware evolution while still covering current threats. Data are delivered through infrastructure that supports streaming, caching, and selective column loading. The dataset size and loading strategy make it practical to use in pedagogical environments—from undergraduate cybersecurity courses to graduate machine learning seminars—on standard student hardware.

\subsection{Motivation and Research Gaps}

The motivation for Binary-30K stems from persistent gaps in the binary analysis research landscape:

\paragraph{Platform Imbalance in Existing Datasets}

Current large-scale binary datasets exhibit severe platform bias. EMBER~\cite{anderson2018ember} and SOREL-20M~\cite{harang2020sorel}—the two most widely used malware detection datasets—consist exclusively of Windows PE files, precluding cross-platform research. Linux-focused datasets exist (e.g., Assemblage~\cite{collberg2024assemblage}), but macOS datasets remain scarce despite 15\% desktop market share and increasing threats~\cite{macos_malware}. Android datasets focus on DEX bytecode and APK structure~\cite{allix2016androzoo,arp2014drebin}, neglecting native ARM libraries. This fragmentation forces researchers to train platform-specific models, limiting generalizability and hindering platform-agnostic detection.

\paragraph{Architectural Diversity for Internet of Things (IoT) Security}

The proliferation of IoT devices and embedded systems has increased the need for binary analysis capabilities beyond x86/x86-64. Botnets such as Mirai and its variants~\cite{antonakakis2017mirai} target routers, cameras, and network equipment running diverse architectures including MIPS, ARM, and PowerPC. More recent IoT malware campaigns have expanded to target exotic architectures: Torii targets x86, ARM, MIPS, Motorola 68k, SuperH, and PowerPC~\cite{avast2018torii}; Okiru specifically targets ARCompact processors used in 1.5 billion IoT devices annually~\cite{fortinet2018okiru}; and 2023 Mirai variants were observed targeting ARM, MIPS, x86, ARC, m68k, and SPARC architectures~\cite{cirt2023mirai}. However, publicly available datasets provide minimal coverage of these exotic architectures, constraining research on cross-architecture malware detection and IoT security.

Binary-30K includes malware samples targeting 15+ distinct architectures, including x86-64, x86, ARM (32/64-bit), MIPS (32/64-bit), RISC-V, ARCompact, m68k, PowerPC, and SuperH. This diversity enables research on architecture-invariant malware detection, cross-architecture transfer learning, and rare architecture recognition—capabilities useful for defending heterogeneous IoT deployments.

\paragraph{Accessibility for Education and Novel Research}

Large-scale datasets like SOREL-20M (\sorelscale{} samples) and EMBER (\emberscale{} samples) often require substantial computational infrastructure—multi-GPU clusters, terabyte-scale storage, and long training runs—which can limit use to well-resourced institutions and hinder adoption in educational contexts. \datasetname takes a different approach: \numrecords{} strategically sampled binaries provide heterogeneous coverage across platform, architecture, and temporal dimensions that would require orders of magnitude more homogeneous samples to match. This intentional scale makes the dataset practical for undergraduate cybersecurity courses, graduate machine learning seminars, and exploratory research on novel architectures or techniques without requiring dedicated compute clusters.

The dataset requires only \compressedsize{} download, fitting comfortably on consumer hardware. Pre-computed BPE tokenization and transformer-ready Arrow/Parquet formats with official splits allow students and researchers to experiment with state-of-the-art deep learning approaches—transformers, state-space models, hybrid architectures—without implementing custom binary preprocessing or navigating complex data loading code. This accessibility supports both pedagogical use in undergraduate and graduate courses and rapid prototyping of novel research ideas where heterogeneous coverage is more valuable than massive scale.

\paragraph{Tokenization and Preprocessing Standardization}

A significant obstacle to reproducibility in binary analysis research is the lack of standardized tokenization. While modern NLP research typically adopts existing tokenizers (GPT-2, LLaMA, tiktoken) to ensure comparability across studies, binary datasets typically distribute raw executables without standardized preprocessing. Researchers consequently develop custom tokenization schemes—byte n-grams, instruction-level tokenization, BPE variants—making it difficult to isolate model innovations from preprocessing effects when comparing published results.

By providing pre-tokenized sequences using the Binary BPE tokenizer family~\cite{bommarito2025binarybpe}, \datasetname enables direct model comparison and lowers the level of binary-format expertise needed to begin experimentation. The Binary BPE tokenizer applies byte-level Byte Pair Encoding specifically designed for executable formats, producing a fixed 65,536-token vocabulary that works across ELF, PE, Mach-O, and APK binaries. This standardized tokenization reduces sequence length compared to byte-level encoding—an important optimization for transformer models with quadratic attention—while preserving the structural information necessary for binary analysis tasks.

\subsection{Contributions}

\datasetname addresses the gaps identified above through several key contributions. First, we provide a \textbf{heterogeneous binary dataset} with three dimensions of diversity: platform (Windows \windowspct{}, Linux \linuxpct{}, macOS \macospct{}, Android \androidpct{}, and other formats \otherpct{}), architecture (15+ CPU architectures including x86, ARM, MIPS, RISC-V, ARCompact), and temporal (2012–2025, spanning compiler evolution and threat landscape changes). This heterogeneous composition, comprising \numrecords{} unique records (100\% SHA-256 deduplicated) at a manageable size (\compressedsize{} download), enables cross-platform transfer learning and architecture-invariant detection research while remaining accessible to academic research groups.

Second, we address the IoT security research gap through \textbf{architectural diversity}: binaries target 15+ CPU architectures, including MIPS, RISC-V, ARCompact, m68k, PowerPC, and SuperH. Our malware selection employs \textbf{stratified sampling} with a platform-first approach that prioritizes platform and architecture diversity through size-based and architecture-based stratification, providing broad coverage across the binary ecosystem.

Third, we deliver a \textbf{deep-learning-ready dataset} where all binaries are pre-tokenized using byte-level BPE (65,536-token vocabulary shared across ELF, PE, Mach-O, and APK formats), packaged in transformer-ready Arrow/Parquet formats, and split into official stratified train/validation/test sets. This infrastructure is purpose-built for transformer architectures: tokenization reduces sequence length for efficient attention, standardized vocabulary enables cross-study comparison, and streaming data loaders integrate directly with PyTorch and HuggingFace. Each binary is accompanied by \textbf{rich metadata} including platform labels, architecture, file format, entropy, size, malware source attribution, and parsed structural features.

Finally, we support accessibility and reproducibility through \textbf{public availability} with documented data loading, and provide \textbf{standardized benchmark tasks} (malware classification, platform identification, architecture recognition) together with official stratified train/validation/test splits distributed via \href{\datasetsplitsurl}{\texttt{mjbommar/binary-30k}} to support future research and comparable evaluations.

\subsection{Paper Organization}

The remainder of this paper is organized as follows. Section~\ref{sec:related} reviews related work in binary analysis datasets and machine learning approaches. Section~\ref{sec:construction} describes the dataset construction methodology, including benign sample collection and the platform-first stratified sampling approach for malware selection. Section~\ref{sec:characteristics} presents comprehensive dataset statistics including platform distribution, architectural diversity, file characteristics, and metadata analysis. Section~\ref{sec:pretokenized} details the BPE tokenization methodology and tokenization statistics. Section~\ref{sec:usecases} discusses research use cases and proposed benchmark tasks. Section~\ref{sec:access} provides practical guidance for accessing and using the dataset, along with licensing information and ethical considerations. Section~\ref{sec:limitations} acknowledges limitations and discusses future work. Section~\ref{sec:conclusion} concludes.

\section{Related Work}
\label{sec:related}

Binary analysis datasets enable machine learning research across malware detection, program analysis, and systems security. We review prior work through three lenses: large malware collections, cross-platform/benign datasets, and public hub datasets, then position \datasetname among them.

\subsection{Malware-Focused Datasets}

The application of machine learning to malware detection has driven the creation of several large-scale datasets, though most remain constrained to Windows executables and specific time periods. These datasets have made significant contributions to the field, yet their platform-specific nature and reliance on engineered features limit their applicability to modern cross-platform security research.

The \textbf{SOREL-20M} dataset~\cite{harang2020sorel} represents one of the largest public malware corpora, providing $\approx$20 million Windows PE samples ($\approx$10M benign, $\approx$10M malicious) collected between 2017 and 2019. The dataset includes rich labels and pre-extracted features that facilitate reproducible research, with approximately 10 million disarmed malware binaries available for download. While the core public release centers on engineered metadata and features, raw binaries can be obtained through the project's referenced sources and partner hosting arrangements, subject to specific terms of use. Despite this availability, the majority of studies leverage SOREL-20M in a feature-based setting rather than working with raw binary content, and the dataset remains exclusively focused on Windows PE files.

The \textbf{EMBER} dataset~\cite{anderson2018ember} has become a cornerstone of reproducible malware detection research, offering 1.1 million Windows PE samples with a balanced 50/50 split between benign and malicious executables. EMBER provides samples as engineered feature vectors of 2,381 dimensions, derived from PE headers, byte histograms, import tables, and string characteristics. While these compact feature representations are convenient for rapid experimentation and have enabled extensive comparative benchmarking across machine learning architectures, they exclude the raw byte sequences necessary for instruction-level or token-level modeling approaches. Like SOREL-20M, EMBER's scope is limited to Windows executables.

The recent \textbf{EMBER2024}~\cite{ember2024} release substantially expands the original EMBER dataset to 3.2 million samples across six file formats, including Win32, Win64, .NET assemblies, Android APKs, Linux ELF binaries, and PDF documents. This broader format coverage enables valuable cross-format comparative studies and reflects the increasingly diverse threat landscape. However, EMBER2024 maintains the feature-only approach of its predecessor, providing no raw byte access for byte-level research. Additionally, despite its expanded scope, the dataset does not include native macOS binaries, leaving an important platform gap in cross-platform malware research.

The \textbf{BODMAS} dataset~\cite{yang2021bodmas} distinguishes itself by providing both raw binaries and extracted features for 57{,}293 malicious and 77{,}142 benign Windows PE samples. Beyond standard feature representations, BODMAS includes curated malware family labels and temporal metadata that support evolution studies and family classification research. These artifacts enable analyses that complement byte-level approaches, while still maintaining the Windows-only focus characteristic of most malware datasets; cross-platform and cross-architecture coverage remain beyond its scope.

Additional malware collections further illustrate the field's Windows-centric orientation. The \textbf{MaleX} dataset~\cite{malex2021} contains approximately 1 million Windows executables with roughly 83\% malware prevalence, while \textbf{MOTIF}~\cite{joyce2023motif} provides 3{,}095 malware samples spanning 454 families with verified labels that enable fine-grained classification studies. Together, these datasets offer substantial scale and family-level supervision useful for establishing Windows classification baselines. However, like their predecessors, these collections remain Windows-centric and typically lack both cross-platform representation and standardized raw-byte tokenization suitable for modern transformer-based approaches.

\subsection{Cross-Platform and Benign Datasets}

\textbf{Assemblage}~\cite{collberg2024assemblage} releases 890K Windows PE and 428K Linux ELF binaries built across 29 compiler/optimization/arch configurations, advancing compilation reproducibility for benign software but excluding malware. Assemblage couples build infrastructure with released binaries to facilitate provenance studies, compiler fingerprinting, and reproducibility analyses.

Android-focused corpora include \textbf{DREBIN}~\cite{arp2014drebin} (5{,}560 apps with static features), \textbf{AndroZoo}~\cite{allix2016androzoo} (millions of APKs), and \textbf{LAMDA}~\cite{lamda2025} (\(>\)1M samples with 369{,}906 malware, 2013–2025). These emphasize APK/DEX artifacts and permissions/features rather than native raw binaries. IoT malware corpora such as \textbf{CUBE-MALIOT-2021}~\cite{cube2021} (47{,}924 samples; ARM/MIPS) support embedded security but remain fragmented.

Linux ELF lacks a public analogue to large Windows malware datasets that combines malware with temporal and architectural breadth. macOS datasets remain scarce (e.g., \textbf{Cyber Science Lab macOS Dataset}~\cite{haddadpajouh2018macos}, 152 samples, 2012–2016), limiting platform-comparative research.

\textbf{Hugging Face datasets.} Public hub datasets illustrate active interest but differ from \datasetname's scope: EMBER 2018 features~\cite{hf_ember2018malware} (\(\sim\)695K PE; feature-only), Malware Samples~\cite{hf_malwaresamples} (raw Windows + CAPEv2 reports, no pre-tokenization), SMU Malware Detection~\cite{hf_smu_malware} (\(\sim\)215K assembly fragments), and Android Malware Dataset~\cite{hf_android_malware} (permissions metadata). These span features, assembly, or metadata rather than multi-platform raw bytes with standardized tokenization.

\subsection{Positioning \datasetname}

\begin{table}[t]
\centering
\footnotesize
\caption{Comparison of Binary-30K with existing binary analysis datasets. To our knowledge, Binary-30K is the \textbf{only public dataset currently available} that simultaneously provides cross-platform coverage (Linux, Windows, macOS, Android), exotic architecture support (MIPS, RISC-V, PowerPC, SH), balanced malware representation (26.93\%), and pre-computed tokenization for transformer models.}
\label{tab:comparison}
\begin{tabular}{@{}lcccccc@{}}
\toprule
\textbf{Dataset} & \textbf{Year} & \textbf{Platform} & \textbf{Samples} & \textbf{Malware\%} & \textbf{Architectures} & \textbf{Pre-tokenized} \\
\midrule
\multicolumn{7}{l}{\textit{Malware-Focused Datasets}} \\
EMBER & 2018 & Windows & 1.1M & 50\% & x86, x86-64 & No \\
EMBER2024 & 2024 & Multi-format & 3.2M & 50\% & x86, x86-64 & No \\
SOREL-20M & 2020 & Windows & 20M & 100\% & x86-64 & No \\
BODMAS & 2019 & Windows & 134K & 57\% & x86, x86-64 & No \\
MaleX & 2023 & Windows & 1M & 83\% & x86, x86-64 & No \\
MOTIF & 2022 & Windows & 3.1K & 100\% & x86, x86-64 & No \\
\midrule
\multicolumn{7}{l}{\textit{Cross-Platform and Benign Datasets}} \\
Assemblage & 2024 & Linux, Windows & 1.3M & 0\% & x86, x86-64, ARM & No \\
DREBIN & 2014 & Android & 5.6K & 50\% & ARM & No \\
AndroZoo & 2016 & Android & Millions & Varies & ARM & No \\
LAMDA & 2025 & Android & 1M+ & 37\% & ARM & No \\
CUBE-MALIOT-2021 & 2021 & IoT/Linux & 47.9K & 100\% & ARM, MIPS & No \\
CS Lab macOS & 2018 & macOS & 152 & 100\% & x86-64 & No \\
ELF Dataset & 2020 & Linux & 3.9K & 50\% & x86-64 & No \\
\midrule
\textbf{Binary-30K} & \textbf{2025} & \textbf{All + Mobile} & \textbf{29,793} & \textbf{26.93\%} & \textbf{All + Exotic} & \textbf{Yes} \\
\bottomrule
\end{tabular}
\end{table}

\textbf{Platform diversity.} To our knowledge, \datasetname is the only public dataset that currently offers malware across Linux, Windows, macOS, and Android in a single ML-ready corpus, complementing benign-only multi-platform datasets (Assemblage) and Windows-focused malware corpora (SOREL-20M, EMBER, BODMAS).

\textbf{Architecture diversity.} Beyond x86/x86-64, \datasetname includes exotic ISAs (e.g., ARM variants, MIPS, PowerPC, RISC-V), enabling IoT and cross-architecture research underrepresented in prior work.

\textbf{Realistic balance.} \datasetname maintains a realistic malware prevalence (\malwareratio), avoiding both extreme imbalance and artificial 50/50 splits.

\textbf{Temporal depth and currency.} \datasetname spans 13 years of binary evolution (2012–2025: benign binaries 2012–2024, malware 2017–2025), capturing both legacy systems (Windows 8, Ubuntu 20.04) and contemporary platforms (Windows 11, Ubuntu 24.04) so that analyses can incorporate older and more recent threats. Malware sources combine historical SOREL-20M Windows samples (2017–2019) with contemporary cross-platform Malware Bazaar samples (2020–2025).

\textbf{Pre-tokenization + raw access.} \datasetname distributes raw executables and standardized byte-level BPE sequences produced by the Binary BPE tokenizer family~\cite{bommarito2025binarybpe}, supporting both raw-byte and transformer-ready workflows—capabilities typically split across feature-only or platform-specific datasets.

\section{Dataset Construction}
\label{sec:construction}

\datasetname was constructed through systematic collection from four primary sources (Linux distribution packages, Windows system binaries, SOREL-20M malware, and Malware Bazaar submissions), each chosen to maximize platform diversity, architectural coverage, and malware representation. This section presents each source independently, emphasizing provenance and collection methodology. Complete extraction procedures, automation scripts, and verification protocols appear in Appendix~\ref{app:collection}.

The final dataset comprises \numunique{} unique binaries (SHA-256 deduplicated), totaling \rawsize{} of raw binary data. The distributed version requires \compressedsize{} compressed download.

\textbf{Dataset size clarification.} Two size figures appear throughout this paper: (1) \textbf{\rawsize{}} represents the sum of all raw binary file sizes in the deduplicated dataset (Table~\ref{tab:platform_dist} ``Size'' column), and (2) \textbf{\compressedsize{}} is the Hugging Face compressed download size using Apache Arrow/Parquet compression with all metadata and tokenization included. The published distribution exhibits zero duplication (100\% SHA-256 deduplicated), simplifying machine learning workflows by reducing duplicate-induced artifacts.

\paragraph{Temporal Coverage Summary}

\datasetname provides temporal depth spanning 13 years (2012–2024 for benign samples; 2017–2025 for malware), reflecting both historical compiler technology and contemporary threat landscapes. Benign binaries derive from: Windows 8 Pro (Build 9200, October 2012), Windows 10 22H2 (Build 19045, October 2022), Windows 11 23H2 (Build 22631, October 2023); Ubuntu 20.04 LTS (April 2020), 22.04 LTS (April 2022), 24.04 LTS (April 2024); Debian 11 Bullseye (August 2021), 12 Bookworm (June 2023); and Alpine Linux 3.18 (May 2023), 3.19 (December 2023). Malware samples span 2017–2025, with SOREL-20M providing historical Windows PE samples (2017–2019) and Malware Bazaar contributing contemporary cross-platform samples (2020–2025). This temporal diversity captures compiler evolution (MSVC 11.0 through MSVC 19.3x, GCC 9.x through GCC 13.x), security feature adoption (ASLR, DEP, CFG, CET, stack canaries, RELRO), and malware technique progression in enterprise environments where legacy systems coexist with modern platforms. Dataset assembly and curation were conducted during 2023–2025.

\subsection{Linux Distribution Packages}
\label{sec:construction:linux}

Linux binaries (\linuxpct{}, \numlinux{} samples) were collected from official distribution images across multiple architectures.

\subsubsection{Collection Methodology}

A custom Docker-based extraction tool~\cite{bommarito2025glaurung} enables cross-architecture collection by copying container filesystems rather than executing commands within containers. Binaries were extracted from standard Unix directories (\texttt{/usr/bin}, \texttt{/bin}, \texttt{/sbin}, \texttt{/usr/local/bin}) with filtering for executable files. The tool automates image pulling, container instantiation, and filesystem extraction across multiple Docker platforms and architectures.

\subsubsection{Distribution Coverage}

The collection spanned four primary Linux distributions, selected to maximize libc diversity (musl vs. glibc), security hardening practices, and architectural coverage:

\textbf{Alpine Linux} (3.18, 3.19) provides musl libc binaries with static linking common in container environments. Collection covered linux/amd64, linux/arm64, and linux/arm/v7 platforms.

\textbf{Ubuntu LTS} (20.04, 22.04, 24.04) represents mainstream glibc-based desktop and server distributions with progressive security hardening (PIE, RELRO, stack canaries) across compiler generations. Collection covered linux/amd64 and linux/arm64 platforms.

\textbf{Debian} (11 Bullseye, 12 Bookworm) provides stable, conservatively-compiled glibc binaries emphasizing long-term support. Collection covered linux/amd64 and linux/arm64 for mainstream architectures. Additionally, Debian's multi-architecture repositories (Debian Ports) provided exotic architecture binaries—MIPS 32/64, PowerPC, RISC-V, m68k, SuperH—contributing approximately \numexotic{} samples critical for IoT and embedded system research.

\textbf{BusyBox} (latest) represents embedded Linux systems through its multi-call binary architecture. Collection covered linux/amd64, linux/arm64, linux/arm/v7, linux/arm/v6, linux/386, linux/ppc64le, and linux/s390x platforms.

Binary selection prioritized functional diversity across utilities, toolchains, network/system services, developer tools, and compression utilities. The collection yielded approximately 11,225 executables across all distributions and architectures before deduplication, resulting in 8,452 unique Linux samples in the final dataset after SHA-256 deduplication. Complete Docker image specifications, extraction scripts, and architecture handling appear in Appendix~\ref{app:collection:linux}.

\subsection{Windows System Binaries}
\label{sec:construction:windows}

Windows binaries (\windowspct{}, \numwindows{} samples) originated from two complementary benign sources providing temporal coverage (2012--2023) and diverse binary types (user-mode executables, kernel-mode drivers).

\subsubsection{Windows ISO Images}

Official Microsoft ISOs provided baseline Windows system binaries across three generations: Windows 8 Pro (Build 9200, October 2012), Windows 10 22H2 (Build 19045, October 2022), and Windows 11 23H2 (Build 22631, October 2023). ISOs were extracted using 7-Zip, with binaries collected from standard directories: \texttt{System32} (64-bit system binaries), \texttt{SysWOW64} (32-bit subsystem on 64-bit Windows), \texttt{Program Files} (64-bit applications), \texttt{Program Files (x86)} (32-bit applications on 64-bit Windows), and \texttt{System32\textbackslash drivers} (kernel-mode drivers).

This tri-generational collection documents compiler evolution and security feature adoption: ASLR and DEP baseline (Windows 8), Control Flow Guard implementation (Windows 10), and modern MSVC toolchain (Windows 11). All binaries originate from digitally signed Microsoft sources, ensuring authenticity. Complete extraction procedures and directory structures appear in Appendix~\ref{app:collection:windows}.

\subsubsection{Windows Update Catalog Drivers}

Device drivers (280 samples) were collected from Microsoft's Windows Update Catalog using an automated browser-based harvesting tool~\cite{bommarito2024wucd}, as the catalog's JavaScript interface prevents traditional HTTP scraping. Driver collection spans network adapters, graphics controllers, storage controllers, USB controllers, and audio devices. These kernel-mode binaries provide insight into kernel-level compilation patterns and security hardening practices absent from user-mode executables.

\subsection{SOREL-20M Malware Collection}
\label{sec:construction:sorel}

To establish a malware baseline and enable malware detection research, we integrated 365 representative samples from SOREL-20M~\cite{harang2020sorel}, a large-scale benchmark dataset for malicious PE detection. SOREL-20M contains \sorelscale{} Windows PE samples (\soreldisarmed{} benign and malicious disarmed binaries available for download) collected 2017--2019 with malware family labels, disassembly metadata, and behavioral features provided by Sophos. The public release centers on engineered metadata and features, while raw binaries can be obtained through the project's referenced sources and partner hosting arrangements, subject to specific terms of use.

Our subset selection prioritized:
\begin{itemize}
	\item \textbf{Malware family diversity}: Ransomware, trojans, backdoors, droppers, rootkits, infostealers, worms
	\item \textbf{Temporal distribution}: Spanning the 2017-2019 collection period
	\item \textbf{Size diversity}: Stratified across file size quartiles to capture varied complexity
	\item \textbf{Packing diversity}: Mix of packed (UPX, Themida, VMProtect) and unpacked samples
\end{itemize}

All SOREL-20M samples are Windows PE executables (PE32 and PE32+), providing ground truth for Windows-targeted malware. SHA-256 deduplication was applied across all dataset sources to prevent duplicates. Binary-30K can be used with \emph{static analysis} of malware samples---no malware execution is necessary, which reduces security risks for researchers. All malware handling followed industry-standard protocols in isolated, network-segregated environments.

\subsection{Malware Bazaar Collection}
\label{sec:construction:malwarebazaar}

Malware Bazaar~\cite{malwarebazaar}, a community-driven malware repository operated by abuse.ch, provided 7,658 cross-platform malware samples selected via platform-first stratified sampling from 20,499 available files. This source complements SOREL-20M's Windows-only coverage by providing macOS malware (568 samples), Android malware (164 samples), and exotic architecture malware (MIPS, RISC-V, PowerPC) targeting IoT and embedded systems. Samples span 2020-2025 collection dates, complementing SOREL-20M's 2017-2019 temporal coverage.

\subsection{Malware Bazaar Sampling}
\label{sec:malware_sampling}

To maximize platform and architecture diversity, we employed platform-first stratified sampling on Malware Bazaar~\cite{malwarebazaar}, selecting 7,658 samples from 20,499 files from randomly sampled archives (37.4\% sampling rate). Combined with 365 SOREL-20M samples, the final dataset includes \nummalware{} malware samples (\malwareratio{} representation), providing balanced class distribution for supervised learning. Table~\ref{tab:malware_sources} presents malware sources by platform.

The sampling strategy prioritized coverage over uniform representation: (1) exhaustive inclusion of all macOS (568) and Android (164) samples to establish mobile and macOS coverage, (2) stratified sampling of Windows (2,500) and Linux (2,000) by file size and architecture, (3) targeted sampling of exotic architectures (MIPS, RISC-V, ARCompact, m68k, SuperH, PowerPC) and diverse formats (scripts, archives, obfuscated binaries). Complete implementation appears in the dataset repository~\cite{bommarito2025datasetpaper}.

\subsection{Quality Assurance}

All binaries underwent multi-stage quality assurance:
\begin{itemize}
	\item \textbf{Source verification}: Benign samples verified against official distribution sources (digitally signed ISOs, package repository signatures). Malware samples verified against SOREL-20M and Malware Bazaar provenance records.
	\item \textbf{Deduplication}: SHA-256 hash-based deduplication across all sources ensures each unique binary appears only once. When identical binaries appeared in multiple sources (e.g., legitimate Windows runtime DLLs from official ISOs that were also bundled with malware samples), we retained only one instance and assigned labels based on provenance. Labels were determined by the original source: Microsoft Visual C++ Runtime API Set forwarder DLLs (\texttt{api-ms-win-crt-*.dll}) extracted from digitally signed Windows ISOs maintain benign labels, as these are legitimate redistributable components frequently bundled by both benign and malicious software. The final dataset contains no duplicate SHA-256 hashes.
	\item \textbf{Provenance and signatures}: Verified source provenance where available (e.g., signed ISOs and package repository signatures). For individual binaries, detected presence of code signatures (Windows Authenticode, macOS code signatures); full chain-of-trust validation of individual binaries was not performed.
	\item \textbf{Format validation}: File format verification using \texttt{file} utility and LIEF parsing. The vast majority of samples parsed successfully. Failures involved corrupted downloads, exotic packers with non-standard headers, or scripts misidentified as binaries. Failed samples were manually inspected and either corrected (re-downloaded) or excluded.
\end{itemize}

Complete QA protocols, deduplication procedures, and verification scripts are available in the dataset repository~\cite{bommarito2025datasetpaper}.

\subsection{Binary Parsing and Metadata Extraction}
\label{sec:construction:parsing}

Metadata extraction employed LIEF (Library to Instrument Executable Formats)~\cite{lief} with format-specific parsers for ELF (Linux/Android/embedded), PE (Windows), Mach-O (macOS), and APK (Android) binaries. Each parser extracts 29 metadata fields across six categories: file identification (SHA-256, MD5, size), platform information (OS family, version, distribution), binary characteristics (format, architecture, type, stripped/packed/signed flags), structural analysis (sections, code/data sizes), dependencies (imports/exports), and complexity metrics (entropy, tokenization). The complete metadata schema appears in Table~\ref{tab:metadata_schema} (Appendix~\ref{app:tok:schema}).

Architecture identification employed a multi-layer fallback strategy: (1) LIEF binary header parsing (ELF \texttt{e\_machine}, PE \texttt{Machine}, Mach-O \texttt{cputype}), (2) Malware Bazaar directory path inference, (3) Debian package filename parsing, and (4) libmagic string pattern matching. This cascading approach achieved 94.29\% architecture identification success across 15+ architectures including common (x86, x86-64, ARM, ARM64) and exotic (MIPS, RISC-V, PowerPC, m68k, SuperH, ARCompact, SPARC) types. Failed LIEF parsing triggered a generic fallback parser that preserved files with diagnostic warnings, with \texttt{parse\_status} and \texttt{parse\_warnings} fields documenting coverage and parsing issues.

The parsing pipeline implementation, format-specific extraction algorithms, robustness strategies, and detailed parsing success statistics are available in the dataset repository~\cite{bommarito2025datasetpaper}.

\section{Dataset Characteristics}
\label{sec:characteristics}

This section summarizes \datasetname's key characteristics—platform diversity, architectural breadth, and malware representation. Comprehensive statistical tables and per-platform analysis appear in Appendix~\ref{app:statistics}.

\subsection{Platform and Malware Distribution}

\begin{table}[t]
\centering
\small
\caption[Platform distribution in Binary-30K dataset]{Platform distribution in Binary-30K dataset (HF distribution), showing record counts, total and average file sizes, and malware representation per platform. The 26.93\% overall malware rate provides solid class balance for machine learning tasks. ``Other'' includes disk images, installer packages, archives, and non-standard formats.}
\label{tab:platform_dist}
\begin{tabular}{lrrrrrr}
\toprule
\textbf{Platform} & \textbf{Count} & \textbf{Percentage} & \textbf{Size (GiB)} & \textbf{Avg (KiB)} & \textbf{Malware} & \textbf{Mal.\%} \\
\midrule
Windows & 17,239 & 57.86\% & 9.95 & 563.6 & 2,719 & 15.77\% \\
Linux & 8,452 & 28.37\% & 1.45 & 167.3 & 2,857 & 33.80\% \\
Other & 3,370 & 11.31\% & 0.49 & 142.4 & 1,715 & 50.89\% \\
macOS\textsuperscript{*} & 568 & 1.91\% & 0.59 & 1020.0 & 568 & 100.0\% \\
Android\textsuperscript{*} & 164 & 0.55\% & 0.70 & 4157.8 & 164 & 100.0\% \\
\midrule
\textbf{Total} & \textbf{29,793} & \textbf{100.0\%} & \textbf{13.18} & \textbf{--} & \textbf{8,023} & \textbf{26.93\%} \\
\bottomrule
\end{tabular}

\vspace{0.5em}
\noindent\small\textsuperscript{*}All macOS and Android samples in this release are malicious (no benign samples). This introduces potential platform-label confounding; see Sections~\ref{sec:characteristics} and~\ref{sec:limitations} for recommended evaluation strategies.
\end{table}

\begin{table}[t]
\centering
\footnotesize
\caption{Malware source breakdown (HF distribution) showing the distribution of malicious samples from SOREL-20M and Malware Bazaar (MB) across platforms.}
\label{tab:malware_sources}
\begin{tabular}{lrrrp{3.2cm}}
\toprule
\textbf{Source} & \textbf{Count} & \textbf{\% Malware} & \textbf{\% Total} & \textbf{Strategy} \\
\midrule
SOREL-20M & 365 & 4.55\% & 1.23\% & Random sampling \\
MB: macOS & 568 & 7.08\% & 1.91\% & All samples \\
MB: Android & 164 & 2.04\% & 0.55\% & All samples \\
MB: Windows & 2,354 & 29.35\% & 7.90\% & Size-stratified \\
MB: Linux & 2,857 & 35.62\% & 9.59\% & Arch-stratified \\
MB: Other & 1,715 & 21.37\% & 5.76\% & Diverse formats \\
\midrule
\textbf{Total} & \textbf{8,023} & \textbf{100.0\%} & \textbf{26.93\%} & \textbf{--} \\
\bottomrule
\end{tabular}
\end{table}

\datasetname comprises \numunique{} unique binaries (SHA-256 deduplicated). The published distribution is Windows-leaning (\windowspct{}) with substantial Linux coverage (\linuxpct{}), and includes macOS (\macospct{}; \nummacos{} samples spanning Intel x86-64, Apple Silicon ARM64, and Universal Mach-O binaries plus 8 macOS-specific disk images and installer packages) and Android (\androidpct{}; \numandroid{} APK samples with native libraries). The remaining \otherpct{} (\numother{} samples) are diverse formats (disk images, installers, archives). This platform distribution enables cross-platform malware detection research and transfer learning experiments beyond Windows-only or Linux-only datasets.

Table~\ref{tab:comparison} positions Binary-30K against existing datasets. To our knowledge, prior datasets do not simultaneously provide macOS malware, Android malware, exotic architectures (MIPS, RISC-V, PowerPC), and pre-computed tokenization. Our balanced platform representation contrasts with SOREL-20M (Windows-only), EMBER (Windows-only), and Assemblage (benign-only), while maintaining manageable scale suitable for machine learning research.

Malware representation totals \nummalware{} samples (\malwareratio{}), avoiding severe class imbalance common in malware datasets (typically $<$1\% or $>$99\%). Dual sourcing from SOREL-20M (365 Windows PE samples collected 2017--2019) and Malware Bazaar (7,658 cross-platform samples) ensures both temporal coverage and platform diversity. Malware Bazaar sampling employed platform-first stratified sampling (Section~\ref{sec:malware_sampling}), including all available macOS and Android malware while maintaining Windows/Linux balance through size- and architecture-stratified sampling. Complete per-platform malware distribution appears in Table~\ref{tab:malware_sources}.

\textbf{Important.} In the current release, all macOS ($n=568$) and Android ($n=164$) platform samples are malicious; there are no benign macOS or Android binaries in \datasetname. This reflects limited availability of redistributable benign binaries for these platforms, but it also introduces a potential label-confounding risk: models trained on the full dataset can, in principle, learn the shortcut ``platform $\rightarrow$ label'' instead of platform-agnostic malware characteristics. Throughout this paper we therefore treat the macOS and Android subsets primarily as held-out targets for cross-platform evaluation and recommend platform-aware splitting strategies for malware classification experiments (see Section~\ref{sec:limitations}).

\subsection{Official Stratified Train/Validation/Test Splits}
\label{sec:official_splits}

To support reproducible machine learning experiments, the Hugging Face release of \datasetname includes official train/validation/test splits (see Section~\ref{sec:access}). These splits are created using a composite stratification key that combines four dimensions for each record: malware status (benign vs.\ malicious), normalized platform (Linux, Windows, macOS, Android, Other), normalized file format (PE, ELF, Mach-O, APK, Other), and an architecture group (common vs.\ exotic). Within each stratum, records are shuffled with a fixed seed (42) and allocated to splits using an approximate 70/15/15 ratio for train/validation/test, with entire very-small strata assigned to training to avoid degenerate splits. Table~\ref{tab:official_splits} summarizes the split sizes and malware prevalence.

\begin{table}[t]
\centering
\small
\caption{Official stratified train/validation/test splits for \datasetname, showing record counts and malware prevalence per split. Splits are created using a composite stratification key over malware status, platform, file format, and architecture group with an approximate 70/15/15 ratio (seed 42).}
\label{tab:official_splits}
\begin{tabular}{lrrrr}
\toprule
\textbf{Split} & \textbf{Samples} & \textbf{Malware} & \textbf{Benign} & \textbf{Mal.\%} \\
\midrule
Train      & 20,849 & 5,613 & 15,236 & 26.92\% \\
Validation & 4,463  & 1,200 & 3,263  & 26.89\% \\
Test       & 4,481  & 1,210 & 3,271  & 27.00\% \\
\midrule
\textbf{Total} & \textbf{\numrecords} & \textbf{\nummalware} & \textbf{21,770} & \textbf{\malwareratio} \\
\bottomrule
\end{tabular}
\end{table}

This procedure preserves the global malware prevalence (\malwareratio{}) while closely matching the overall platform, file format, and architecture distributions in each split. The official splits therefore provide a standard configuration for benchmarking malware classification, platform identification, and architecture recognition tasks while retaining the cross-platform and cross-architecture diversity of the full dataset. As noted above, the underlying macOS and Android subsets remain malware-only across all splits, so recommended evaluation strategies in Section~\ref{sec:limitations} continue to apply. Detailed stratification statistics, including the full set of strata and chi-square verification tests, are available in the public dataset documentation and project repository.

\subsection{Architectural Diversity}

\begin{figure}[t]
\centering
\includegraphics[width=0.8\columnwidth]{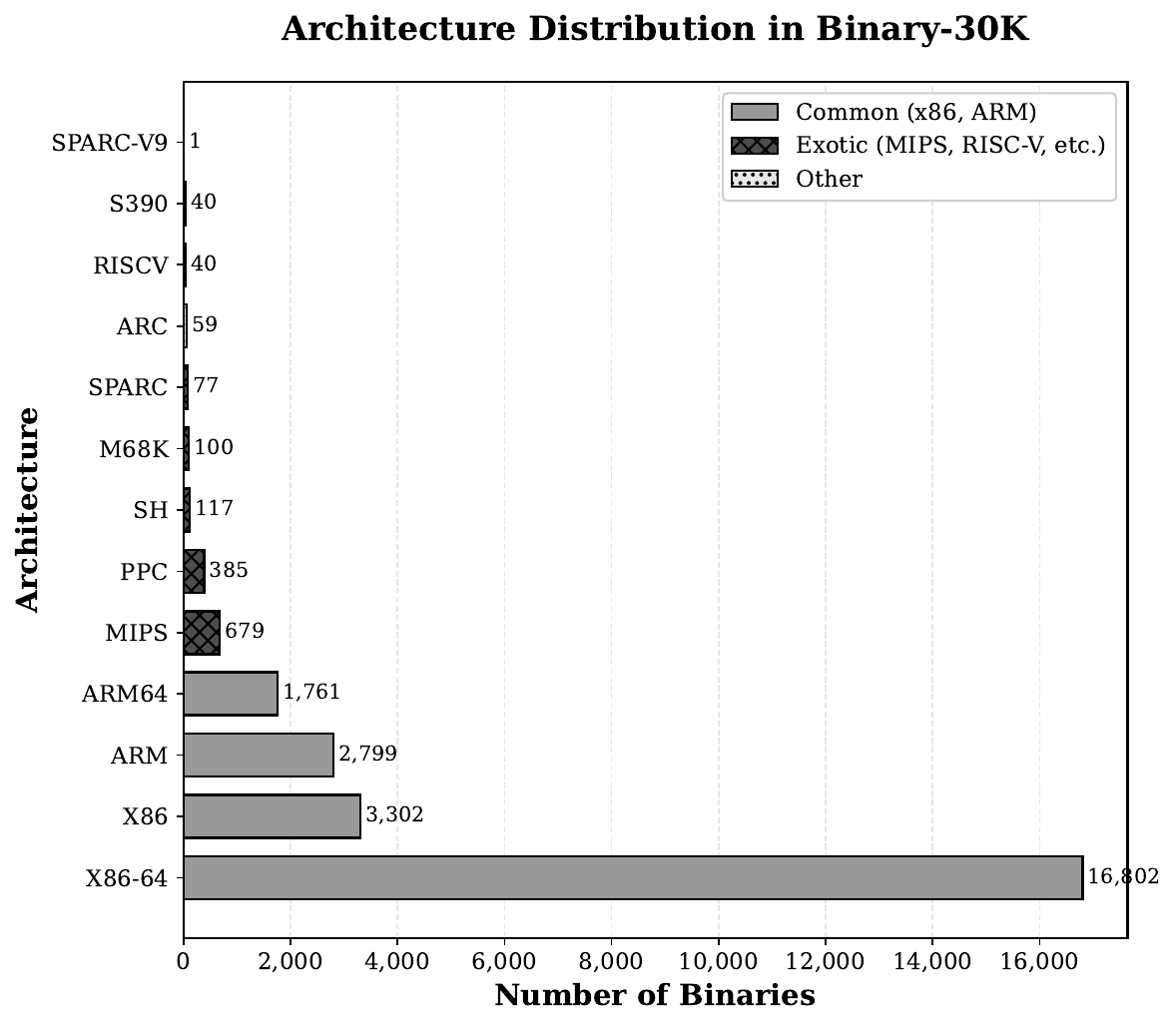}
\caption{Architecture distribution in the Binary-30K dataset (excluding Unknown and Not-Applicable). The dataset includes comprehensive coverage of common architectures: x86-64 (16,802 samples, 56.40\%), x86-32 (3,302 samples, 11.08\%), ARM (2,799 samples, 9.39\%), and ARM64 (1,761 samples, 5.91\%) shown in orange. Exotic architectures shown in green include MIPS (679 samples, 2.28\%), PowerPC (385 samples, 1.29\%), SH (117 samples), m68k (100 samples), SPARC (77 samples and 1 SPARC-v9), RISC-V (40 samples), ARCompact (59 samples), and s390 (40 samples), totaling 1,498 exotic architecture samples. This fills a significant gap in existing binary analysis datasets and enables research on architecture-specific malware and cross-architecture analysis.}
\label{fig:architecture}
\end{figure}

Binary-30K spans 15+ distinct CPU architectures. Automated architecture identification succeeded for 94.29\% of binary files (26,162 binaries out of 27,744 applicable samples, excluding \numnotapplicable{} scripts and text files). The distribution reflects both market prevalence and intentional oversampling of exotic architectures for IoT security research. Common architectures dominate: x86-64 (16,802 samples, 56.40\%), x86 (3,302 samples, 11.08\%), ARM (2,799 samples, 9.39\%), and ARM64 (1,761 samples, 5.91\%). Exotic architectures critical for IoT research include: MIPS (679 samples, 2.28\%), PowerPC (385 samples, 1.29\%), RISC-V (40 samples, 0.13\%), m68k (100 samples, 0.34\%), SuperH (117 samples, 0.39\%), ARCompact (59 samples, 0.20\%), SPARC (77 samples, 0.26\%), and S/390 (40 samples, 0.13\%).

This architectural diversity helps address a limitation in existing datasets: most malware datasets provide x86/x86-64 only, precluding research on cross-architecture detection, architecture fingerprinting, or IoT-specific threat modeling. The approximately 6,050 non-x86 architecture samples enable development of architecture-invariant malware signatures and study of architecture-specific attack patterns, particularly relevant for IoT botnets like Mirai variants that specifically target MIPS and ARM devices.

Architecture identification uses a two-stage approach: (1) LIEF binary parsing extracts machine type from executable headers (ELF e\_machine field, PE Machine field, Mach-O cputype), and (2) for Malware Bazaar samples, folder structure provides fallback identification when parsing fails or returns ambiguous results. A small portion of samples (5.31\%, \numunknownarch{} samples) remain architecturally unidentified (Unknown), while 6.88\% (\numnotapplicable{} samples) are marked as not-applicable for scripts, text files, disk images, and multi-architecture containers like APKs that lack a single definable architecture. Investigation of the unknown architecture samples revealed that approximately 60\% are ELF binaries with valid but unmapped machine types, and 4\% are PE malware with deliberately corrupted headers (\texttt{machine\_type=0x0000})---a real-world anti-analysis technique. The dramatic improvement in architecture identification from earlier dataset versions (previously 33\% unknown) results from enhanced LIEF parser support for exotic architectures and improved fallback mechanisms. These unidentified and not-applicable samples retain value for format-agnostic ML techniques and improving automatic architecture detection. Complete architecture breakdown with per-architecture statistics appears in Appendix~\ref{app:stats:architecture}.

\subsection{File Characteristics}

\begin{table}[t]
\centering
\small
\caption{File format distribution (HF distribution) after re-identification using libmagic and bucketing. Categories include native binaries (PE, ELF, Mach-O), mobile packages (APK), text, scripts, and common containers (ZIP/JAR/DMG/TAR/PKG). Unknown represents formats not identified by LIEF or libmagic.}
\label{tab:file_format}
\begin{tabular}{lrr}
\toprule
\textbf{Format} & \textbf{Count} & \textbf{Percentage} \\
\midrule
PE (PE32/PE32+) & 17,184 & 57.68\% \\
ELF (ELF32/ELF64) & 8,452 & 28.37\% \\
Unknown & 1,708 & 5.73\% \\
TEXT & 1,208 & 4.05\% \\
Mach-O & 560 & 1.88\% \\
SCRIPT & 374 & 1.26\% \\
APK & 164 & 0.55\% \\
Other (ZIP/JAR/DMG/TAR/PKG/misc) & 77 & 0.26\% \\
Archive (ar) & 65 & 0.22\% \\
DATA & 1 & 0.00\% \\
\midrule
\textbf{Total} & \textbf{29,793} & \textbf{100.00\%} \\
\bottomrule
\end{tabular}
\end{table}

\begin{figure*}[t]
\centering
\includegraphics[width=\textwidth]{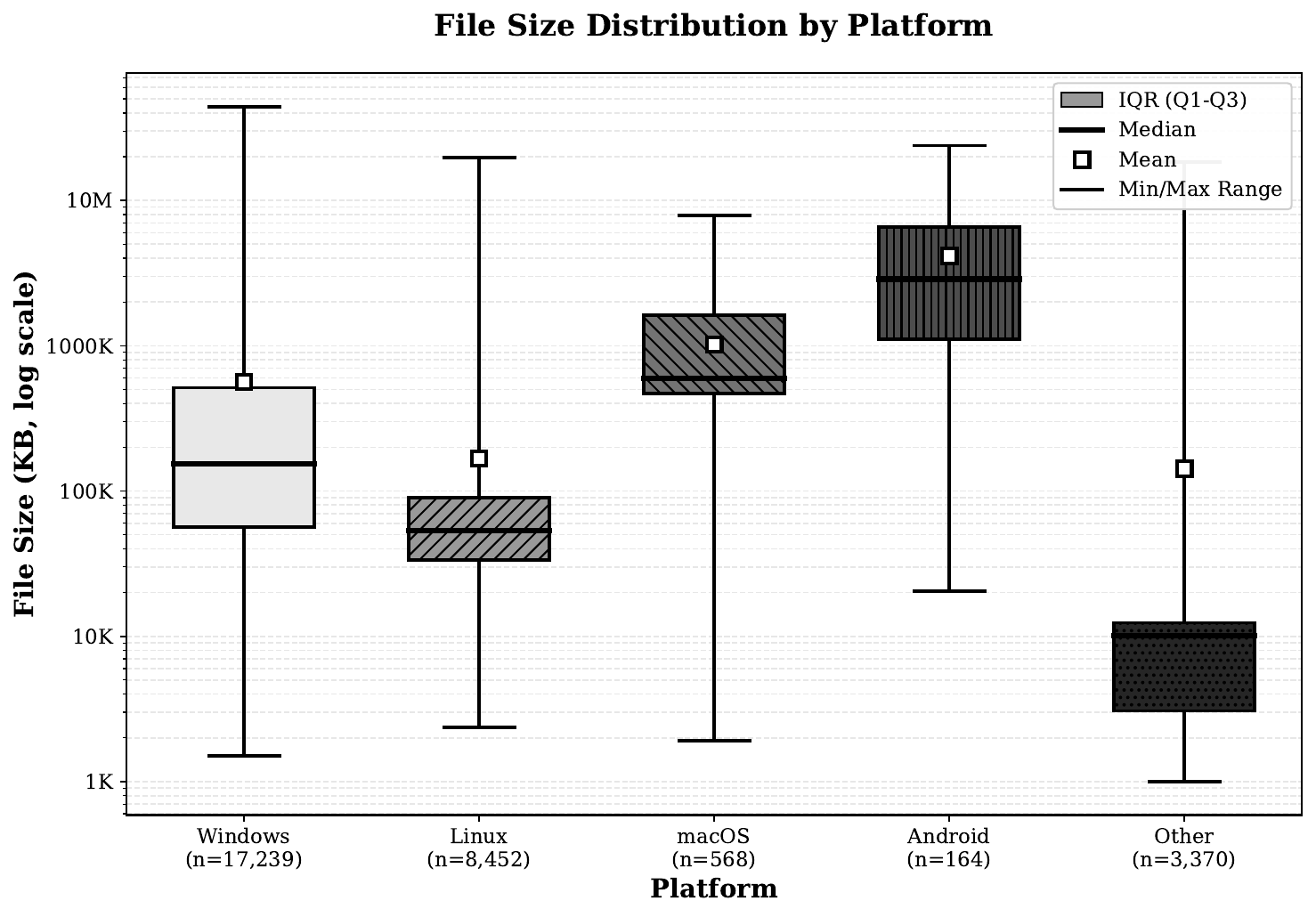}
\caption{File size distribution by platform with log-scale x-axis. Each subplot shows the interquartile range (IQR, shaded blue region) and median (red dashed line) for a specific platform. Android binaries are significantly larger (median: 2.87 MB) due to APK packaging, while Linux binaries are smallest (median: 53.5 KB) due to the prevalence of small utility programs. The wide range of file sizes within each platform (spanning 3-4 orders of magnitude) demonstrates the diversity of binary types included in the dataset, from small embedded firmware to large application bundles.}
\label{fig:file_size}
\end{figure*}

\begin{table}[t]
\centering
\small
\caption{Entropy statistics comparing benign and malware binaries. Higher entropy (closer to 8.0) indicates greater randomness, often associated with packing, encryption, or compression. Mean values shown with standard deviation in parentheses. High entropy threshold: $\geq$7.0.}
\label{tab:entropy}
\begin{tabular}{lrrrrrrrc}
\toprule
\textbf{Category} & \textbf{Count} & \textbf{Min} & \textbf{Q1} & \textbf{Median} & \textbf{Q3} & \textbf{Max} & \textbf{Mean (SD)} & \textbf{High ($\geq$7.0)} \\
\midrule
Benign & 21,770 & 0.25 & 4.87 & 5.68 & 6.14 & 8.00 & 5.34 (1.17) & 299 (1.4\%) \\
Malware & 8,023 & 0.00 & 5.70 & 6.42 & 7.84 & 8.00 & 6.50 (1.23) & 2,959 (36.9\%) \\
\bottomrule
\end{tabular}
\end{table}

The dataset encompasses diverse binary formats reflective of modern computing ecosystems. Primary categories include PE (57.68\%, 17,184 Windows executables), ELF (28.37\%, 8,452 Linux/Unix binaries), text artifacts (4.05\%, 1,208 files such as configuration fragments and assembly source), script files (1.26\%, 374 Python/Perl/shell scripts), Mach-O binaries (1.88\%, 560 macOS samples including Universal builds), and APK packages (0.55\%, 164 Android applications). Note that the 568 macOS platform samples comprise 560 Mach-O binaries plus 8 additional macOS-specific formats (DMG disk images and PKG installers). Unknown formats comprise 5.73\% (1,708 samples), capturing genuinely unclassified or corrupted binaries, exotic archive formats, and obfuscated malware that resisted identification by both LIEF and libmagic. Complete format distribution with per-format analysis appears in Appendix~\ref{app:stats:format}.

File sizes range from small utilities to large application bundles, with substantial variation across platforms. Platform-specific patterns emerge: Linux binaries tend to be smaller (median: 53.5 KB, reflecting the Unix philosophy of specialized utilities), Windows binaries exhibit moderate sizes (median: 153.3 KB, reflecting monolithic executable patterns), macOS binaries tend to be substantially larger (median: 597.3 KB, due to code signing, Universal binary overhead, and bundled frameworks), and Android APKs have the largest sizes (median: 2.87 MB, reflecting compressed application bundles with resources and native libraries). The diverse file size distribution reflects differences across platforms and formats and is useful for studying size-invariant detection techniques. Detailed per-platform size statistics appear in Appendix~\ref{app:stats:size}.

Shannon entropy (0.0--8.0 scale) reveals binary complexity and compression characteristics (Table~\ref{tab:entropy}). Malware samples exhibit significantly higher median entropy (6.42 bits/byte) than benign samples (5.68 bits/byte)---a 13\% elevation reflecting widespread obfuscation practices. More dramatically, 36.9\% of malware samples (2,959 binaries) exceed the high-entropy threshold of 7.0 bits/byte compared to only 1.4\% of benign samples (299 binaries), representing a 27-fold difference. This concentration of high-entropy malware reflects prevalent use of packing (UPX, custom packers), encryption, and compression to evade signature-based detection. The dataset's balanced representation of both normal-entropy and high-entropy malware provides strong discriminative signals for detection research while capturing real-world evasion techniques. Per-platform entropy analysis appears in Appendix~\ref{app:stats:entropy}.

\subsection{Dataset Integrity}

The published Hugging Face dataset applies SHA-256 deduplication, yielding \numunique{} unique binaries from \numrecords{} total records with zero duplicate records in the final distribution. This deduplication was applied across all sources (benign Linux/Windows/macOS/Android + SOREL-20M + Malware Bazaar) to ensure dataset quality and prevent train/test contamination. Note that earlier dataset versions contained BusyBox multi-call binaries and system utility hardlinks with higher duplication rates; these were removed in the final curation to provide a deduplicated distribution suitable for machine learning workflows. Malware samples from SOREL-20M and Malware Bazaar contribute unique binaries without overlap. Complete dataset construction details appear in Section~\ref{sec:construction}.

Metadata extraction via LIEF achieved parsing success across diverse formats, with libmagic-based reidentification providing fallback classification. Overall, 5.73\% of samples (1,708 binaries) remain with unknown format designation. These unknown format samples include corrupted files, unsupported exotic formats (e.g., OCaml libraries, G-IR databases), misclassified assembly source and text fragments, and obfuscated malware with deliberately malformed headers. Despite unknown format designations, these samples contribute to dataset diversity and reflect that real-world binary collections include parsing challenges and unusual formats.

\section{Pre-tokenized Version}
\label{sec:pretokenized}

\datasetname includes pre-computed byte-level BPE tokenization for transformer-based research, reducing preprocessing effort and supporting standardized model comparison. For comprehensive analysis of the Binary BPE tokenizer family---including training methodology, vocabulary characteristics, learned patterns, and compression performance across platforms---see the companion tokenizer paper~\cite{bommarito2025binarybpe}. This section provides key usage information; complete training parameters, metadata schema, and per-platform details appear in Appendix~\ref{app:tokenization}.

\subsection{Tokenization Overview}

We provide pre-computed byte-level BPE tokenization using the 64K-vocabulary instance of the Binary BPE tokenizer family~\cite{bommarito2025binarybpe}.\footnote{Binary BPE tokenizer available at \href{\tokenizerurl}{\texttt{mjbommar/binary-tokenizer-001-64k}}} This platform-agnostic tokenizer, trained on Binary-30K with a 65,536-token vocabulary, achieves approximately 3.37× compression versus byte-level encoding (3.37 bytes/token weighted average across platforms), reducing sequence lengths for efficient transformer processing. The vocabulary includes 256 single-byte tokens plus 65,280 learned multi-byte patterns. Table~\ref{tab:tokenization} summarizes key statistics; sequence lengths range from hundreds to over 6,000,000 tokens, with median token counts of 19,861 (Linux), 49,823 (Windows), 336,476 (macOS), and 1,454,293 (Android).

\begin{table}[t]
\centering
\small
\caption{Tokenization statistics by platform using byte-pair encoding (BPE) with a 65{,}536-token vocabulary (HF distribution). Compression ratio is bytes per token (higher = better compression; e.g., 3.48 means each token represents 3.48 bytes on average).}
\label{tab:tokenization}
\begin{tabular}{lrrrr}
\toprule
\textbf{Platform} & \textbf{Avg Tokens} & \textbf{Comp. Ratio} & \textbf{Avg Unique} & \textbf{Median Tokens} \\
\midrule
Windows & 258,886 & 3.48 & 12,850 & 49,823 \\
Linux & 76,880 & 3.34 & 6,134 & 19,861 \\
macOS & 384,649 & 2.84 & 16,147 & 336,476 \\
Android & 2,445,585 & 2.00 & 28,133 & 1,454,293 \\
Other & 71,770 & 3.05 & 1,511 & 2,686 \\
\midrule
\textbf{Overall} & \textbf{331,556} & \textbf{2.89} & \textbf{--} & \textbf{--} \\
\bottomrule
\end{tabular}
\end{table}

\subsection{Benefits of Pre-Tokenized Data}

The pre-tokenized format provides several advantages for deep learning research:

\begin{enumerate}
\item \textbf{Transformer-based malware detection:} Researchers can directly load pre-tokenized sequences into transformer models (BERT, GPT, etc.) without implementing custom tokenization pipelines. This standardization facilitates model comparison and reproducibility across studies.

\item \textbf{Representation learning:} Pre-tokenized data enables training of binary embeddings using self-supervised learning objectives (masked language modeling, contrastive learning). These embeddings can transfer to downstream tasks such as malware classification, similarity search, and code clone detection.

\item \textbf{Sequence modeling research:} The pre-computed token sequences support research on efficient sequence modeling techniques for long contexts, including state-space models (Mamba~\cite{gu2023mamba}, Mamba-2~\cite{dao2024mamba2}) and hybrid architectures (Jamba~\cite{lieber2024jamba}). These models achieve linear-time complexity $O(n)$ while maintaining competitive performance with transformers. Binary analysis presents unique challenges due to extreme sequence lengths (up to 3,000,000 tokens for large Android APKs), making this dataset a useful testbed for evaluating long-range architectures.

\item \textbf{Cross-platform analysis:} The platform-agnostic tokenization enables direct comparison of binaries across different platforms and architectures, supporting transfer learning and domain adaptation research. Models trained on Linux binaries can be evaluated on Windows or Android samples without re-tokenization.
\end{enumerate}

\subsection{Handling Long Sequences}

\begin{itemize}
\item \textbf{Sliding windows:} Overlapping chunks with aggregation; simple and compatible with pre-trained models.
\item \textbf{Hierarchical chunking:} Encode sections (code, data, imports) then aggregate; preserves structure.
\item \textbf{Long-range models:} Use state-space models (Mamba~\cite{gu2023mamba}, Mamba-2~\cite{dao2024mamba2}) or hybrid architectures (Jamba~\cite{lieber2024jamba}) to process million-token sequences more efficiently than standard transformers.
\end{itemize}

For malware detection baselines, 512-token sliding windows offer a good balance of coverage and cost. Additional strategies include hierarchical processing of independent chunks and state-space architectures (Mamba, Mamba-2, Jamba) for efficient million-token sequence handling.

\subsection{Data Format}

The pre-tokenized version uses Apache Arrow format with 31 metadata fields (excluding tokens) plus tokenization outputs (tokens, token\_count, compression\_ratio, unique\_tokens). This format provides columnar storage with efficient compression, enabling memory-mapped loading for streaming workflows and selective column access. Complete schema documentation appears in Table~\ref{tab:metadata_schema} (Appendix~\ref{app:tok:schema}); loading examples and access instructions appear in Section~\ref{sec:access}.

\section{Use Cases}
\label{sec:usecases}

{\datasetname}'s multi-platform coverage, malware representation, and pre-tokenized format support applications across binary analysis, machine learning, and cybersecurity research and education. This section highlights practical use cases.

\subsection{Malware Detection and Classification}

The dataset's \malwareratio{} malware ratio (\nummalware{} malware samples) supports supervised malware detection without extreme resampling. Researchers can train classifiers using:

\begin{itemize}
	\item \textbf{Byte-level CNNs:} Convolutional neural networks operating on raw bytes can learn local patterns indicative of malicious behavior, such as shellcode, packing signatures, and obfuscation techniques~\cite{raff2017malware}.

	\item \textbf{Transformer-based models:} Pre-tokenized sequences support training transformer models on binary content to capture long-range dependencies and contextual patterns.

	\item \textbf{Hybrid architectures:} Combine metadata features (size, entropy, imports) with learned representations from raw bytes.
\end{itemize}

Cross-platform malware detection is particularly valuable: models trained on one platform can be evaluated on others (Linux, Windows, macOS, Android) to quantify domain gaps and identify platform-agnostic signatures.

\subsection{Cross-Platform Binary Analysis}

{\datasetname}'s coverage (57.86\% Windows, 28.37\% Linux, 1.91\% macOS, 0.55\% Android) supports research on cross-platform binary understanding:

\begin{itemize}
	\item \textbf{Format-agnostic models:} Researchers can train models that operate on raw bytes without format-specific preprocessing, enabling unified analysis of ELF, PE, Mach-O, and APK formats. This approach is particularly valuable for detecting format polyglots and multi-platform malware campaigns.

	\item \textbf{Platform classification:} As a baseline task, platform classification (predicting whether a binary targets Linux, Windows, macOS, or Android) tests a model's ability to learn platform-specific structural patterns. This task complements malware detection and can serve as a pre-training objective.

	\item \textbf{Architecture recognition:} With 15+ distinct architectures represented, the dataset enables research on architecture identification from raw bytes. This capability is valuable for firmware analysis, IoT security, and reverse engineering workflows where architecture metadata may be missing or unreliable.

	\item \textbf{Cross-compilation detection:} Comparing binaries compiled from the same source code for different platforms enables research on compiler fingerprinting and build reproducibility.
\end{itemize}

\subsection{Representation Learning and Embeddings}

The dataset's size (\numrecords{} records) and diversity support representation learning research:

\begin{itemize}
	\item \textbf{Self-supervised pre-training:} Masked language modeling (MLM) on the pre-tokenized sequences enables learning binary embeddings without requiring labels. Models pre-trained on Binary-30K can transfer to downstream tasks such as malware family classification, vulnerability detection, and code similarity.

	\item \textbf{Contrastive learning:} Pairs of binaries (same source code compiled for different platforms, different versions of the same software, etc.) enable contrastive learning objectives that learn platform-invariant and version-invariant representations.

	\item \textbf{Binary similarity search:} Learned embeddings enable semantic similarity search across binaries (e.g., malware variant detection, clone identification).

	\item \textbf{Dimensionality reduction visualization:} Projecting high-dimensional binary embeddings to 2D/3D using t-SNE or UMAP enables visualization of the binary landscape, revealing clusters corresponding to platforms, compilers, and malware families.
\end{itemize}

\subsection{IoT and Embedded Systems Security}

{\datasetname}'s exotic architecture coverage (MIPS, RISC-V, ARCompact, m68k, SuperH, PowerPC) supports IoT security research, a setting often underrepresented in existing public datasets:

\begin{itemize}
	\item \textbf{IoT malware detection:} The dataset includes malware targeting routers (MIPS), industrial control systems (ARCompact), and embedded devices (m68k, SuperH). Researchers can develop specialized detection models for these architectures or study cross-architecture transfer learning.

	\item \textbf{Firmware analysis:} Many embedded system binaries lack standard headers and metadata, requiring models that operate on raw bytes. Binary-30K's diverse format coverage supports research on format-agnostic firmware analysis.

	\item \textbf{Botnet family classification:} Linux malware in the dataset includes Mirai variants and other botnet malware targeting IoT devices. Fine-grained malware family classification can inform threat intelligence and incident response.
\end{itemize}

\subsection{Mobile Platform Security}

Android malware (164 APK samples) enables mobile security research:

\begin{itemize}
	\item \textbf{APK structure analysis:} Android packages combine DEX bytecode, native ARM/ARM64 libraries (.so files), and resources. Multi-modal models can analyze these components jointly or separately.

	\item \textbf{Native library malware detection:} Many Android malware samples include obfuscated native libraries to evade DEX-based detection. Binary-30K supports research on native library analysis complementing existing DEX-focused datasets~\cite{allix2016androzoo,arp2014drebin}.

	\item \textbf{Cross-platform mobile threats:} Comparing Android malware in \datasetname with iOS or desktop malware from other sources can reveal cross-platform attack patterns and shared codebase indicators.
\end{itemize}

\subsection{Model Efficiency and Compression}

Binary analysis presents extreme sequence lengths (up to 3M tokens), motivating research on efficient model architectures:

\begin{itemize}
	\item \textbf{Long-context models:} Binary-30K provides a testbed for evaluating state-space models (Mamba~\cite{gu2023mamba}, Mamba-2~\cite{dao2024mamba2}) and hybrid transformer-Mamba architectures (Jamba~\cite{lieber2024jamba}) that aim to handle sequences up to 1M+ tokens more efficiently than standard transformers. Unlike many NLP datasets, binary sequences often lack natural segmentation, requiring models to handle continuous byte streams.

	\item \textbf{Hierarchical models:} Chunk-based or hierarchical models that first encode local regions and then aggregate global context may offer efficiency advantages over flat transformers.

	\item \textbf{Pruning and quantization:} Researchers can evaluate model compression techniques on this dataset for deployment on resource-constrained environments such as endpoint security agents and embedded systems.
\end{itemize}

\subsection{Educational Applications}

Beyond research, {\datasetname}'s size and distribution format make it well-suited for educational settings:

\begin{itemize}
	\item \textbf{Cybersecurity courses:} Instructors can integrate Binary-30K into malware analysis, reverse engineering, and security fundamentals courses. The dataset provides real-world malware samples in a controlled format, reducing the need for students to handle raw malware files or navigate complex binary tooling.

	\item \textbf{Machine learning pedagogy:} The dataset serves as a domain-specific ML benchmark for courses on deep learning, sequence modeling, and applied AI. Binary classification provides a practical alternative to commonly used image/text datasets, exposing students to real-world challenges such as extreme sequence lengths, class imbalance, and domain adaptation.

	\item \textbf{Hands-on labs and assignments:} Manageable dataset size (\compressedsize{} download) enables classroom use on standard student laptops without specialized hardware. Memory-mapped access allows efficient use without requiring the full dataset in RAM. Simple loading utilities reduce installation complexity and allow students to focus more on model development than data engineering.

	\item \textbf{Cross-disciplinary learning:} The dataset bridges computer security and machine learning, supporting courses at the intersection of these fields and enabling students to apply modern AI techniques to cybersecurity problems.
\end{itemize}

Instructors adopting \datasetname in coursework should also consult Appendix~\ref{app:educator_safety} for recommended safe-handling practices and institutional considerations when working with malware samples in teaching environments.

\subsection{Benchmark Tasks}

To facilitate research and enable standardized evaluation, we propose several benchmark tasks:

\begin{enumerate}
	\item \textbf{Binary malware classification:} Binary classification (benign vs. malware) with train/validation/test splits stratified by platform and malware source. Evaluation metrics: F1-score, AUC-ROC, false positive rate at 95\% true positive rate.

	\item \textbf{Platform identification:} Multi-class classification (Linux, Windows, macOS, Android, Other). Evaluation metric: Accuracy, confusion matrix analysis.

	\item \textbf{Architecture recognition:} Multi-class classification across 15+ architectures. Handles class imbalance via macro-averaged F1.

	\item \textbf{Malware family attribution:} Fine-grained malware family classification using external labels from VirusTotal or Malpedia. Evaluation metric: Macro F1, top-k accuracy.
\end{enumerate}

\textbf{Note:} Detailed benchmark task results, including baseline model performance, reference training/evaluation scripts, and pretrained BERT-like encoder models, will be presented in companion work. This paper focuses on dataset construction, characteristics, and methodology. Official stratified train/validation/test splits suitable for these tasks are already provided via the Hugging Face release at \href{\datasetsplitsurl}{\texttt{mjbommar/binary-30k}} (see Section~\ref{sec:official_splits}).

\section{Dataset Access and Licensing}
\label{sec:access}

\datasetname is publicly available through the Hugging Face Datasets hub~\cite{bommarito2025binary30k} with permissive licensing for research and educational use. This section provides complete access instructions, licensing information, ethical guidelines, and support resources.

\subsection{Access}

\paragraph{Primary Access via Hugging Face Datasets}

The recommended access method uses the Hugging Face \texttt{datasets} library, which handles automatic downloading, caching, and memory-mapped loading. The primary entry point is the stratified-splits dataset at \href{\datasetsplitsurl}{\texttt{mjbommar/binary-30k}}, which exposes train/validation/test splits suitable for benchmarking. The original unsplit, fully tokenized dataset remains available at \href{\datasettokenizedurl}{\texttt{mjbommar/binary-30k-tokenized}} for users who wish to construct custom splits or perform low-level analyses. Complete loading examples, API reference, and quick-start code appear in the dataset's Hugging Face cards and in Appendix~\ref{app:tok:schema}.

\paragraph{Quick Start Example}

Loading and exploring the dataset with streaming (no download required):

\begin{lstlisting}[language=Python, basicstyle=\small\ttfamily,
  frame=single, numbers=left, numberstyle=\tiny,
  breaklines=true, showstringspaces=false,
  commentstyle=\color{gray}, keywordstyle=\bfseries]
from datasets import load_dataset

# Option 1: Streaming mode (instant access, no download)
ds = load_dataset("mjbommar/binary-30k", split="train",
                  streaming=True)
first = next(iter(ds))
print(first["platform"], first["architecture"])

# Option 2: Download full splits for offline use
ds = load_dataset("mjbommar/binary-30k")
train_ds, val_ds, test_ds = ds["train"], ds["validation"], ds["test"]
\end{lstlisting}

The same API can be used to load the original, unsplit, fully tokenized dataset~\cite{bommarito2025binary30ktokenized}. Complete examples and tutorials are available in the dataset repository~\cite{bommarito2025datasetpaper}.

\paragraph{Alternative Access}

For air-gapped environments or users preferring direct downloads, the datasets are available via the Hugging Face Hub web interface or command-line tools. The dataset includes 31 metadata fields (excluding tokens, 32 including tokens), enabling efficient memory-mapped loading and streaming access. The tokenization outputs include: tokens, token\_count, compression\_ratio, and unique\_tokens. Complete field descriptions and schema documentation appear in Appendix~\ref{app:tok:schema}.

\subsection{License}

\paragraph{Dataset Compilation License:}

The \textbf{\datasetname compilation}—including the curated collection, metadata, structural analysis, pre-computed tokenization, and dataset organization—is released under \textbf{Creative Commons Attribution 4.0 International (CC-BY-4.0)}. This license covers our contributed work: sample selection methodology, metadata extraction, tokenization, and dataset documentation.

\paragraph{Component Binary Licenses:}

Individual binaries retain their original source licenses and terms of use. Users are responsible for compliance with all applicable licenses:

\begin{itemize}
\item \textbf{Linux packages}: Various open-source licenses (GPL, MIT, Apache, BSD, etc.) per distribution policies
\item \textbf{Windows binaries}: Microsoft Software License Terms and Windows Update EULA
\item \textbf{SOREL-20M malware}: Sophos AI research license terms~\cite{harang2020sorel}
\item \textbf{Malware Bazaar samples}: abuse.ch terms of use~\cite{malwarebazaar}
\end{itemize}

Users should \textbf{cite original sources} when using specific subsets (e.g., cite SOREL-20M when using those malware samples) and \textbf{not redistribute} as complete operating systems or commercial software collections. When in doubt, consult source dataset documentation and licensing terms.

Complete licensing terms, component-specific licenses, redistribution guidelines, and legal compliance requirements are available in the dataset repository.

\subsection{Ethical Use and Safety}

\datasetname contains \nummalware{} active malware samples and is intended for defensive security research and education. Users are responsible for complying with institutional policies and handling malware safely.

\paragraph{Safe handling.}
Work with raw binaries requires isolated environments: dedicated VMs with no host filesystem access, disabled network connectivity, and snapshot/restore workflows. For machine learning research using only pre-tokenized data, the tokenized dataset eliminates exposure to raw binaries. ML researchers should verify with their institutions whether tokenized malware representations require the same handling protocols as raw binaries.

\paragraph{Institutional compliance.}
Researchers and instructors must obtain appropriate institutional approvals before using the dataset, which may include IT security clearance, IRB review, or risk assessments. Policies vary by organization and jurisdiction. Appendix~\ref{app:educator_safety} provides guidance for educators.

\paragraph{Dual-use considerations.}
This dataset supports defensive applications (malware detection, architecture recognition, security education). Users must not employ the dataset for offensive operations, unauthorized access, or malware creation. When publishing research involving adversarial techniques or evasion, clearly describe defensive benefits and follow responsible disclosure norms.

Additional safety guidelines and liability disclaimers are available in the dataset repository~\cite{bommarito2025datasetpaper}.

\subsection{Support}

For technical support, bug reports, or research inquiries, use the Hugging Face Discussions forum:

\begin{itemize}
\item \textbf{Hugging Face Discussions}: \href{https://huggingface.co/datasets/mjbommar/binary-30k/discussions}{Dataset discussion forum}
\item \textbf{Documentation}: This paper provides comprehensive dataset documentation
\item \textbf{Dataset card}: Hugging Face dataset card includes quick-start examples and schema reference
\end{itemize}

Usage best practices, evaluation metrics, train/test split recommendations, community contribution guidelines, and contact information are available in the dataset repository~\cite{bommarito2025datasetpaper}.

\section{Limitations and Future Work}
\label{sec:limitations}

While \datasetname provides cross-platform malware coverage, several important limitations should be acknowledged and may motivate future enhancements.

\textbf{Scale.}
\datasetname contains \numunique{} unique binaries, which is substantially smaller than SOREL-20M's $\approx$20 million samples. The dataset emphasizes platform and architecture diversity, comprehensive metadata (31 fields), and pre-computed tokenization for transformer models—features that are not simultaneously provided by many larger, platform-specific datasets. When considering multi-platform scope and feature richness, \datasetname is more directly comparable to EMBER (1.1M samples), a widely adopted single-platform benchmark. In practice, we view \datasetname primarily as a research and educational resource and as a complement to larger, platform-specific corpora rather than a drop-in replacement for industry-scale telemetry.

\textbf{Architecture Coverage.}
x86-64 binaries dominate the dataset (56.40\%), reflecting the current prevalence of x86-64 across Linux, Windows, and macOS ecosystems. ARM architectures (ARM32, ARM64) comprise approximately 15\% of the dataset, primarily from Android and embedded Linux samples. RISC-V and other exotic architectures represent approximately 3\% of samples. As IoT devices proliferate and ARM-based systems become more prominent in security research, future iterations should prioritize expanded ARM coverage and inclusion of emerging architectures.

\textbf{Platform Gaps.}
iOS is notably absent from \datasetname, due to Apple's restrictions on binary distribution and the scarcity of publicly available iOS malware samples. Similarly, other mobile platforms (KaiOS, HarmonyOS, Tizen) lack representation. Additionally, no firmware images or bootloaders are included. Future work should expand mobile platform coverage as legitimate research sources and ethical distribution mechanisms become available, potentially including iOS, proprietary mobile OSes, and firmware-level binaries.

\textbf{Temporal Heterogeneity and Potential Biases.}
Binary-30K spans 13 years of binary evolution (2012–2025: benign samples 2012–2024; malware 2017–2025), which provides valuable temporal depth but introduces heterogeneity that may affect machine learning research. Benign binaries represent different compilation eras: Windows 8 binaries (October 2012) use MSVC 11.0 with early ASLR/DEP implementations, Ubuntu 20.04 binaries (April 2020) use GCC 9.x with mature stack protector and RELRO hardening, while Windows 11 (October 2023) and Ubuntu 24.04 (April 2024) binaries represent modern compiler technology with Control Flow Guard, Intel CET, and advanced link-time optimization. Malware samples exhibit similar temporal spread: SOREL-20M Windows PE samples (2017–2019) reflect historical packing techniques and evasion methods, while Malware Bazaar cross-platform samples (2020–2025) include contemporary obfuscation, multi-stage loaders, and platform-specific exploits.

This temporal diversity reflects realistic enterprise environments where legacy systems (Windows 8 end-of-life January 2023; Ubuntu 20.04 LTS supported until April 2025) coexist with modern platforms, but it introduces potential confounds for machine learning models. Models may inadvertently learn era-specific compiler artifacts (e.g., MSVC 11.0 vs. MSVC 19.3x code generation patterns, GCC 9.x vs. GCC 13.x optimization strategies) rather than malware-specific behavioral patterns. Similarly, malware detection models trained on historical SOREL-20M samples (2017–2019) may underperform on contemporary threats that employ newer evasion techniques, while models trained on recent Malware Bazaar samples may not recognize legacy malware variants.

Researchers studying temporal robustness, concept drift, or time-invariant malware signatures should consider temporal stratification when constructing train/test splits. The official stratified splits (Section~\ref{sec:official_splits}) stratify by platform, file format, architecture, and malware status but do not explicitly stratify by compilation date or malware collection year, as precise compilation timestamps are not uniformly available across all sources. Future dataset releases should incorporate explicit temporal metadata (compilation timestamps, malware first-seen dates) to enable controlled temporal experiments and longitudinal studies of compiler evolution, malware technique progression, and model degradation over time.

\textbf{Malware Diversity and Sources.}
All malware samples originate from two sources: SOREL-20M (365 samples, 2017–2019) and Malware Bazaar (7,658 samples, 2020–2025). This limited sourcing may introduce geographic or linguistic biases in malware families. Additionally, platform-level labels are provided, but individual malware family or campaign labels are not included; adding fine-grained family annotations would enhance utility for family-specific detection and attribution tasks. Future expansion should incorporate additional malware corpora and provide family-level annotations from sources such as YARA rules or VirusTotal classification data.

\textbf{Platform Imbalance and Label Confounding.}
While \datasetname provides balanced Linux/Windows representation (\linuxpct{} vs \windowspct{}) with healthy malware/benign distributions on both platforms, the current version has an important limitation: as discussed in Section~\ref{sec:characteristics} and shown in Table~\ref{tab:platform_dist}, all macOS ($n=568$) and all Android ($n=164$) samples are malicious (100\% malware). This reflects the limited availability of publicly shareable benign binaries for these platforms rather than a deliberate design choice. This creates a potential label leakage issue where classifiers could learn the spurious association ``platform $\rightarrow$ label'' rather than meaningful binary characteristics.

We are actively working to address this limitation in future releases by acquiring benign macOS and Android binaries from vetted sources. In the interim, we recommend researchers using \datasetname for malware classification tasks adopt one of three evaluation strategies: (1) train models on Linux and Windows samples only, then evaluate cross-platform generalization to macOS/Android separately; (2) use platform-stratified splits that prevent models from exploiting platform metadata; or (3) exclude explicit platform features during training to encourage learning of platform-agnostic patterns. The official stratified train/validation/test splits released via \href{\datasetsplitsurl}{\texttt{mjbommar/binary-30k}} (Section~\ref{sec:official_splits}) preserve the underlying macOS/Android imbalance and are primarily intended for Linux/Windows malware classification and cross-platform evaluation; they do not in themselves resolve the absence of benign macOS and Android samples.

\textbf{Static Analysis Only.}
\datasetname focuses exclusively on static features extracted with Python tooling such as LIEF and libmagic-based classifiers. Dynamic execution traces—including system calls, network communications, file operations, and heap allocations—are not included. While static analysis is valuable for model pretraining and efficient detection, dynamic traces provide complementary behavioral signals. Future work should integrate multimodal datasets combining static binary features with dynamic execution profiles, enabling models to learn both structural and behavioral patterns.

\textbf{File Type Focus.}
The dataset includes only executable binaries and shared libraries (ELF, PE, Mach-O). Scripts (Python, Bash, PowerShell), bytecode (Java, .NET), WebAssembly, and firmware images are excluded. As interpreted languages and WebAssembly-based malware grow in importance, future iterations should expand to include these file types and their corresponding tokenization schemes.

\textbf{Benchmark Task Results.}
While Section~\ref{sec:usecases} proposes several benchmark tasks for standardized evaluation (malware classification, platform identification, architecture recognition, and malware family attribution), detailed benchmark results including baseline model performance and end-to-end training/evaluation scripts are not included in this paper. Official stratified train/validation/test splits suitable for these tasks are already available via the Hugging Face release (Section~\ref{sec:official_splits}). Companion work will present baseline results along with BERT-like encoder models pretrained on the dataset, enabling researchers to leverage transfer learning for downstream binary analysis tasks.

\textbf{Future Work.}
We plan to release regular updates incorporating newly discovered malware samples and expanding platform/architecture coverage, with the expanded dataset likely to be released as Binary-50K. We welcome community contributions, including additional binaries, metadata enrichments, and evaluation benchmarks. Future priorities include: (1) publishing benchmark task results with baseline performance along with pretrained BERT-like encoder models in companion work, (2) adding malware family and campaign labels, (3) integrating dynamic analysis traces, (4) expanding to interpreted languages and firmware, and (5) providing time-series snapshots to enable longitudinal studies of binary and malware evolution. These enhancements would further strengthen \datasetname as a resource for binary security research.

\section{Conclusion}
\label{sec:conclusion}

\datasetname is the first heterogeneous binary dataset specifically designed for deep learning in malware detection and binary analysis. Spanning Windows, Linux, macOS, and Android platforms; 15+ CPU architectures including IoT-critical exotic architectures (MIPS, RISC-V, ARCompact, m68k, PowerPC, SuperH); and a 2012–2025 temporal range capturing compiler evolution and malware landscape changes, the dataset provides the heterogeneity necessary for cross-platform transfer learning and architecture-invariant detection research. Pre-computed BPE tokenization with a 65,536-token vocabulary trained on binary data makes the dataset immediately usable with transformer, state-space, and hybrid architectures through PyTorch and HuggingFace integration, eliminating the preprocessing barriers that have hindered deep learning adoption in binary security research. With \numunique{} unique binaries and \malwareratio{} malware representation achieved through platform-first stratified sampling, Binary-30K delivers both heterogeneous coverage and balanced class distribution.

\subsection{Key Contributions and Strengths}

Binary-30K makes four primary contributions to binary analysis research. First, it provides \textbf{heterogeneous coverage} across platforms (Windows, Linux, macOS, Android), architectures (15+), and temporal range (2012–2025)—no other publicly available dataset combines macOS malware, Android malware, and exotic architecture coverage in a single resource. Second, it addresses \textbf{IoT and embedded security research} through stratified sampling that prioritizes exotic architectures (MIPS, RISC-V, ARCompact, m68k, PowerPC, SuperH), providing the first substantial public collection of malware targeting these platforms. Third, it delivers \textbf{production-ready tokenization} through pre-computed BPE with a 65,536-token vocabulary, enabling immediate use with transformers, state-space models, and hybrid architectures without custom preprocessing. Fourth, it ensures \textbf{accessibility and reproducibility} at \compressedsize{} compressed with official stratified train/validation/test splits and standardized benchmark tasks, making the dataset practical for educational environments and rapid research prototyping.

\subsection{Impact on Binary Security Research}

The dataset's heterogeneous composition and deep-learning-ready infrastructure enable research on cross-platform transfer learning, architecture-invariant detection for IoT security, novel deep learning architectures (long-context transformers, Mamba, Jamba), and reproducible educational use. The standardized tokenization and comprehensive documentation lower barriers for both undergraduate instruction and graduate research, while modest computational requirements enable rapid prototyping of novel techniques.

\subsection{Limitations and Future Directions}

As detailed in Section~\ref{sec:limitations}, key areas for future work include fine-grained malware family labels, dynamic analysis traces, temporal expansion (Binary-50K), and iOS coverage. Official stratified splits are already available via Hugging Face (Section~\ref{sec:official_splits}), with baseline benchmarks and pretrained encoder models to be presented in companion work.

\subsection{Call to Action and Community Engagement}

Binary-30K is released under CC-BY-4.0 licensing (for dataset compilation, metadata, and tokenization) to support broad adoption and derivative works. The underlying binary files retain their original source licenses. Complete access instructions, licensing details, and ethical guidelines appear in Section~\ref{sec:access}. We invite the research community to:

\begin{itemize}
	\item \textbf{Publish baseline results}: Establish benchmark performance on canonical tasks (malware detection accuracy, cross-platform transfer learning, architecture classification) to enable rigorous comparison across future research
	\item \textbf{Contribute annotations}: Cross-reference samples with threat intelligence sources to add malware family labels, campaign attributions, and behavioral tags
	\item \textbf{Propose dataset extensions}: Suggest additional data sources, platforms, or architectures that would broaden research utility
	\item \textbf{Share derived datasets}: Create and publish filtered subsets, augmented versions, or task-specific variants to support specialized research communities
	\item \textbf{Report issues and corrections}: Identify mislabeled samples, parsing errors, or metadata inconsistencies to improve dataset quality
\end{itemize}

Binary-30K—the first publicly available heterogeneous binary dataset designed for deep learning—provides a common foundation for the next generation of sequence-based, cross-platform, architecture-aware malware detection systems. As adversaries increasingly deploy sophisticated malware across diverse platforms and architectures, defenders need detection capabilities that generalize beyond single platforms and adapt rapidly to emerging threats. We believe the architectural innovations driving breakthroughs in natural language processing and computer vision—transformers, state-space models, and transfer learning—can fundamentally improve defensive security when applied to binary analysis.

Our hope is that Binary-30K enables the research community to pursue this vision. The dataset supports developing detection systems that protect users across Windows, Linux, macOS, Android, and IoT devices, training models accessible to under-resourced organizations, and establishing whether deep learning can provide defenders with the speed, accuracy, and adaptability needed to counter modern malware campaigns at scale.

\bibliographystyle{plainnat}
\bibliography{bibtex/references}

\clearpage
\appendix
\section{Platform Collection Details}
\label{app:collection}

This appendix summarizes the binary collection methodology for each platform. Complete reproduction procedures, Docker commands, package lists, and extraction scripts are available in the \href{\datasetrepourl}{dataset's GitHub repository}.

\subsection{Linux Collection}
\label{app:collection:linux}

Linux binaries (\linuxpct{}, \numlinux{} samples) were collected from eight distributions: Alpine (3.18, 3.19), Debian (11, 12), Ubuntu LTS (20.04, 22.04, 24.04), and BusyBox (1.37.0), using Docker containers and official package managers. Debian Ports repositories contributed exotic architecture binaries (MIPS, PowerPC, RISC-V, m68k, SuperH) to the dataset's \numexotic{} total exotic architecture samples (Section~\ref{sec:characteristics}). Complete package lists and Docker commands are available on GitHub.

\subsection{Windows Collection}
\label{app:collection:windows}

Windows binaries (\windowspct{}, \numwindows{} samples post-deduplication) were collected from two benign sources: official Microsoft ISOs extracted 16,149 binaries (Windows 8 Pro Build 9200, Windows 10 22H2 Build 19045, Windows 11 23H2 Build 22631), and Windows Update Catalog provided 280 device drivers via automated browser harvesting~\cite{bommarito2024wucd}. After SHA-256 deduplication across all dataset sources and addition of Windows malware (2,719 samples from SOREL-20M and Malware Bazaar), the final dataset contains \numwindows{} unique Windows binaries. All benign binaries originate from digitally signed Microsoft sources.

\subsection{macOS Collection}
\label{app:collection:macos}

macOS binaries (\macospct{}, 568 samples) were collected exclusively from Malware Bazaar, representing all available macOS malware at collection time. No benign macOS samples were included due to Apple software redistribution restrictions. The collection spans three architectural variants: Intel x86-64 Mach-O binaries (pre-2020), Apple Silicon ARM64 Mach-O binaries (2020+), and Universal (Fat) binaries containing both architectures. Universal binaries enable research on multi-architecture packaging and cross-platform code sharing.

\subsection{Android Collection}
\label{app:collection:android}

Android binaries (\androidpct{}, 164 samples) were collected from Malware Bazaar APK files, with native libraries (.so files) extracted from the \texttt{lib/} directory. No benign Android samples were included due to Google Play Store licensing restrictions. Extracted ELF shared objects target ARM architectures (32-bit ARMv7, 64-bit ARMv8) and contain JNI integration for Java/Kotlin interoperation. The collection represents major Android malware categories: banking trojans, spyware, adware, ransomware, and botnet clients.

\subsection{Verification Procedures}

All binaries underwent file format validation (LIEF parsing), architecture detection, SHA-256 deduplication, and metadata extraction. Benign sources were verified via signed ISOs and package repository signatures. Complete verification scripts and detailed procedures available in the \href{\datasetrepourl}{project repository}.

\section{Extended Statistics}
\label{app:statistics}

This appendix provides supplementary statistical tables and detailed breakdowns that complement Section~\ref{sec:characteristics}. Complete statistical analysis scripts and raw data files are available in the \href{\datasetrepourl}{project repository}.

\subsection{Architecture Statistics}
\label{app:stats:architecture}

Table~\ref{tab:architecture} presents the complete architecture distribution across Binary-30K, revealing both mainstream and specialized instruction set architectures (ISAs).

\begin{table}[t]
\centering
\small
\caption{Architecture distribution in Binary-30K dataset (HF distribution). Unknown indicates architecture could not be determined from binary metadata. Not-applicable includes scripts and text files without architecture.}
\label{tab:architecture}
\begin{tabular}{lrrp{5.5cm}}
\toprule
\textbf{Architecture} & \textbf{Count} & \textbf{Percentage} & \textbf{Primary Platforms} \\
\midrule
x86-64 & 16,802 & 56.40\% & Linux, Windows, macOS \\
x86 & 3,302 & 11.08\% & Linux, Windows \\
ARM (32-bit) & 2,799 & 9.39\% & Android, Linux \\
ARM64 & 1,761 & 5.91\% & macOS, Android, Linux \\
Not-applicable & 2,049 & 6.88\% & Scripts, text files \\
Unknown & 1,582 & 5.31\% & Mixed \\
MIPS & 679 & 2.28\% & Linux \\
PowerPC & 385 & 1.29\% & Linux \\
SH & 117 & 0.39\% & Linux \\
m68k & 100 & 0.34\% & Linux \\
SPARC & 77 & 0.26\% & Linux \\
ARC & 59 & 0.20\% & Linux \\
RISC-V & 40 & 0.13\% & Linux \\
S390 & 40 & 0.13\% & Linux \\
SPARC-v9 & 1 & 0.00\% & Linux \\
\midrule
\textbf{Total} & \textbf{29,793} & \textbf{100.00\%} & \textbf{--} \\
\bottomrule
\end{tabular}
\end{table}

Binary-30K provides architecture coverage useful for developing cross-architecture binary analysis techniques. The four mainstream architectures (x86-64, x86-32, ARM32, ARM64) collectively represent approximately 83\% of the dataset. Beyond mainstream architectures, the dataset includes approximately 1,500 samples from exotic and specialized architectures prevalent in IoT devices, embedded systems, network equipment, and specialized computing applications.

\textbf{MIPS} (679 samples, 2.28\%) represents the most substantial exotic architecture coverage. MIPS processors are common in network equipment (routers, switches), embedded systems, and IoT devices. Historically, major IoT botnets (Mirai and variants) have specifically targeted MIPS-based devices, making MIPS coverage essential for IoT malware research. The dataset includes both 32-bit and 64-bit MIPS variants, as well as big-endian and little-endian configurations.

\textbf{PowerPC} (385 samples, 1.29\%) encompasses binaries for the PowerPC architecture used historically in Apple Macintosh computers (pre-Intel transition), IBM servers, and various embedded systems. While desktop PowerPC is largely obsolete, the architecture remains in use in specialized computing, automotive systems, and industrial applications.

\textbf{SuperH} (117 samples, 0.39\%) covers the SuperH RISC architecture developed by Hitachi and used in embedded systems, automotive electronics, and consumer devices.

\textbf{Additional exotic architectures} include RISC-V (40 samples, open-source ISA gaining traction in research and embedded systems), ARCompact/ARC (59 samples, Synopsys ARC architecture for deeply embedded applications), m68k (100 samples, Motorola 68000 legacy architecture), SPARC (78 samples total including SPARC and SPARC-v9, Sun Microsystems architecture), and S390 (40 samples, IBM mainframe architecture). These exotic architectures collectively total 1,498 samples (5.03\%), representing the long tail of computing platforms.

\subsubsection{Unknown Architecture}

A portion of samples (5.31\%, \numunknownarch{} samples) have unknown architecture identification. This reflects the diversity of file formats and parsing challenges inherent in large-scale binary collection:

\begin{enumerate}
\item \textbf{Non-executable formats}: Scripts, archives (ZIP, TAR, DMG), disk images (ISO, PKG), and installers that contain executables but are not themselves executable binaries with architecture metadata
\item \textbf{Format complexity}: Multi-architecture containers (Universal binaries, APKs with multiple native libraries) where architecture must be inferred from contents rather than container headers
\item \textbf{Obfuscated binaries (4\%, 374 PE samples)}: Investigation revealed deliberate header corruption as an anti-analysis technique. All 374 unknown PE files have \texttt{machine\_type=0x0000} (invalid value), evading static analysis tools that validate PE headers. These files execute on Windows (permissive loader) but fail strict validation, demonstrating real-world evasion tactics
\item \textbf{LIEF limitations}: Expanded architecture mapping in parsing library supports 26+ ELF machine types (see \texttt{src/parser/elf\_parser.py}), including common (x86, x86-64, ARM, ARM64, MIPS, PowerPC, RISC-V) and exotic architectures (SuperH, m68k, SPARC, ARCompact, Xtensa, and others), though some ELF machine types remain unmapped to avoid false positives on corrupted headers
\item \textbf{Metadata stripped}: Binaries with removed or corrupted headers where architecture information is absent or unreliable
\end{enumerate}

Architecture identification uses a two-stage pipeline: (1) LIEF binary parsing extracts machine type from executable headers (ELF \texttt{e\_machine}, PE Machine field, Mach-O cputype), yielding 94.29\% identification rate (26,162 identified out of 27,744 applicable files, excluding \numnotapplicable{} not-applicable scripts and text files), and (2) for Malware Bazaar samples, folder structure provides fallback identification (\texttt{malware-bazaar/<platform>/<format>/<arch>/}).

\subsection{File Format Distribution}
\label{app:stats:format}

Table~\ref{tab:file_format} presents the file format distribution across Binary-30K, revealing diverse binary formats reflective of modern computing ecosystems. Each format embodies distinct structural characteristics, analysis challenges, and platform-specific conventions.

PE format represents 57.68\% of the dataset (17,184 samples), encompassing Windows executables and dynamic libraries in both PE32 (32-bit) and PE32+ (64-bit) variants. PE binaries demonstrate average file size of 563.6 KiB with substantial variation and average entropy of 5.70.

ELF format represents 28.37\% of the dataset (8,452 samples), serving as the standard executable format for Linux and Unix systems in both ELF32 and ELF64 variants. ELF binaries demonstrate average file size of 167.3 KiB (notably smaller than PE binaries) and average entropy of 5.65 (comparable to PE).

Mach-O format represents 1.88\% of the dataset (560 samples), serving as the executable format for macOS systems. Mach-O encompasses both single-architecture binaries (x86-64, ARM64) and Universal (Fat) binaries containing code for multiple architectures. Mach-O binaries exhibit average file size of 1,020.0 KiB (substantially larger than Linux/Windows due to code signing data, entitlements, and bundled frameworks) and average entropy of 5.51.

APK format represents 0.55\% of the dataset (164 samples), serving as the Android application package format. APK files are ZIP archives containing DEX bytecode, native libraries (.so files), resources, and manifests. APK files demonstrate dramatically larger average file size (4,157.8 KiB) due to bundled resources and average entropy of 7.01 (highest among major formats due to compression).

Unknown formats represent 5.73\% of the dataset (1,708 samples) after reclassification with libmagic. Note that common categories have been reclassified: text files (4.05\%, 1,208 samples), script files (1.26\%, 374 samples), and archive files (0.22\%, 65 samples) are now tracked separately. The remaining Unknown category includes corrupted or partial files, obfuscated malware with deliberately malformed headers, exotic formats not recognized by standard parsing tools, and miscellaneous binary data files.

\subsection{Entropy Distribution}
\label{app:stats:entropy}

Shannon entropy serves as a fundamental complexity measure, quantifying the randomness or information density of binary content (see Table~\ref{tab:entropy} in Section~\ref{sec:characteristics}). The dataset exhibits median entropy of 5.89 with moderate variation (Q1: 5.02, Q3: 6.44), indicating typical compiled code characteristics. Malware samples exhibit significantly higher median entropy (6.42) than benign samples (5.68), reflecting widespread use of packing, compression, and encryption for detection evasion. Among malware samples, 36.9\% exhibit high entropy ($\geq$7.0), far exceeding the 1.4\% rate observed in benign binaries.

\subsection{File Size Statistics}
\label{app:stats:size}

Table~\ref{tab:size_stats} presents comprehensive file size statistics stratified by platform.

\begin{table}[t]
\centering
\small
\caption{File size statistics by platform (HF distribution). Android APKs are largest (median 2,864 KiB), followed by macOS binaries (median 597 KiB). Linux and Windows executables show compact distributions, while Other files are smallest (median 10 KiB).}
\label{tab:size_stats}
\begin{tabular}{lrrrrrrr}
\toprule
\textbf{Platform} & \textbf{Count} & \textbf{Min (KiB)} & \textbf{Q1 (KiB)} & \textbf{Med (KiB)} & \textbf{Q3 (KiB)} & \textbf{Max (MiB)} & \textbf{Mean (KiB)} \\
\midrule
Windows & 17,239 & 1.5 & 56.3 & 153.3 & 513.2 & 44.18 & 563.6 \\
Linux & 8,452 & 2.4 & 33.6 & 53.5 & 89.8 & 19.78 & 167.3 \\
macOS & 568 & 1.9 & 468.1 & 597.3 & 1624.5 & 7.85 & 1020.0 \\
Android & 164 & 20.4 & 1108.1 & 2870.8 & 6546.7 & 23.85 & 4157.8 \\
Other & 3,370 & 1.0 & 3.1 & 10.1 & 12.4 & 18.35 & 142.4 \\
\midrule
\textbf{Overall} & \textbf{29,793} & \textbf{1.0} & \textbf{33.6} & \textbf{--} & \textbf{--} & \textbf{44.18} & \textbf{--} \\
\bottomrule
\end{tabular}
\end{table}

File size distributions reveal substantial variation across platforms reflecting diverse application types and platform conventions. Linux binaries demonstrate compact median size of 53.5 KiB (Q1: 33.6 KiB, Q3: 89.8 KiB) reflecting the Unix philosophy of specialized utilities. Windows binaries demonstrate median size of 153.3 KiB (Q1: 56.3 KiB, Q3: 513.2 KiB) with larger footprint due to richer headers and resources. macOS binaries demonstrate median size of 597.3 KiB (Q1: 468.1 KiB, Q3: 1,624.5 KiB) reflecting code signing data, Universal binary overhead, and bundled frameworks. Android APKs demonstrate largest median size of 2,870.8 KiB (Q1: 1,108.1 KiB, Q3: 6,546.7 KiB) reflecting bundled DEX bytecode, native libraries, and compressed resources.

\section{Tokenization and Metadata Details}
\label{app:tokenization}

This appendix provides technical details on tokenization methodology and metadata schema. Complete implementation details, field examples, and loading code appear in the \href{\datasetrepourl}{dataset repository}.

\subsection{Tokenizer Training Parameters}
\label{app:tok:training}

The Binary-30K tokenizer (\texttt{mjbommar/binary-tokenizer-001-64k}) is an instance of the Binary BPE tokenizer family~\cite{bommarito2025binarybpe}. It employs Byte Pair Encoding (BPE) with a 65,536-token vocabulary trained on a superset of this dataset. Vocabulary utilization is 67.2\% (44,058 of 65,536 tokens used across the dataset), indicating good coverage without saturation. The full tokenizer training pipeline, scaling studies across 4K--64K vocabularies, and additional diagnostics are described in the companion tokenizer paper and its \href{\binarybpepaperurl}{open-source implementation}.

Common learned patterns include ELF/PE/Mach-O headers, x86-64 instruction prefixes, null byte sequences, and ASCII string fragments (\texttt{/lib/}, \texttt{.so}, \texttt{.dll}).

\subsection{Complete Metadata Schema}
\label{app:tok:schema}

Binary-30K provides 31 metadata fields per sample (32 including \texttt{tokens}), extracted via LIEF library and custom parsers. Table~\ref{tab:metadata_schema} summarizes the complete schema. Detailed field descriptions and examples appear in Section~\ref{sec:construction:parsing}.

\begin{table}[ht]
	\centering
	\scriptsize
	\setlength{\tabcolsep}{3pt}
	\caption{Complete metadata schema with field descriptions and data types}
	\label{tab:metadata_schema}
	\begin{tabular}{p{3.3cm}p{1.8cm}p{7.8cm}}
		\toprule
		\textbf{Field Name}         & \textbf{Type} & \textbf{Description}                                                                                                   \\
		\midrule
		\multicolumn{3}{l}{\textbf{File Identification}}                                                                                                                     \\
		\texttt{sha256}             & string        & SHA-256 hash (64 hex chars) for deduplication                                                                          \\
		\texttt{md5}                & string        & MD5 hash (32 hex chars) for legacy compatibility                                                                       \\
		\texttt{file\_size}         & integer       & File size in bytes                                                                                                     \\
		\texttt{file\_path}         & string        & Original file path in collection source                                                                                \\
		\texttt{file\_name}         & string        & Base filename without directory path                                                                                   \\
		\texttt{file\_id}           & string        & Unique identifier within dataset                                                                                       \\
		\\
		\multicolumn{3}{l}{\textbf{Platform Information}}                                                                                                                    \\
		\texttt{platform}           & string        & Platform: \texttt{linux}, \texttt{windows}, \texttt{macos}, \texttt{android}, \texttt{other}                           \\
		\texttt{os\_family}         & string        & OS family: \texttt{debian}, \texttt{ubuntu}, \texttt{alpine}, \texttt{windows}, \texttt{macos}, etc.                   \\
		\texttt{os\_version}        & string        & Specific OS version (e.g., ``22.04'', ``Windows 10'') or ``unknown''                                                   \\
		\texttt{distribution}       & string        & Linux distribution name if applicable (e.g., ``Ubuntu'', ``Debian'')                                                   \\
		\\
		\multicolumn{3}{l}{\textbf{Binary Characteristics}}                                                                                                                  \\
		\texttt{file\_format}       & string        & Format: \texttt{ELF32}, \texttt{ELF64}, \texttt{PE32}, \texttt{PE32+}, \texttt{Mach-O}, \texttt{APK}, \texttt{unknown} \\
		\texttt{architecture}       & string        & Architecture: \texttt{x86-64}, \texttt{ARM64}, \texttt{ARM32}, \texttt{MIPS}, \texttt{RISC-V}, etc.                    \\
		\texttt{binary\_type}       & string        & Type: \texttt{executable}, \texttt{library}, \texttt{driver}, \texttt{object}                                          \\
		\texttt{is\_stripped}       & boolean       & Debug symbols removed? (true/false)                                                                                    \\
		\texttt{is\_packed}         & boolean       & Packing/compression detected? (true/false)                                                                             \\
		\texttt{is\_signed}         & boolean       & Code signature present? (true/false)                                                                                   \\
		\\
		\multicolumn{3}{l}{\textbf{Structural Analysis}}                                                                                                                     \\
		\texttt{num\_sections}      & integer       & Number of sections in binary (ELF/PE/Mach-O)                                                                           \\
		\texttt{code\_size}         & integer       & Total size of executable code sections (bytes)                                                                         \\
		\texttt{data\_size}         & integer       & Total size of data sections (bytes)                                                                                    \\
		\texttt{sections}           & list          & List of section names and properties (JSON array)                                                                      \\
		\\
		\multicolumn{3}{l}{\textbf{Dependencies}}                                                                                                                            \\
		\texttt{num\_imports}       & integer       & Number of imported functions                                                                                           \\
		\texttt{num\_exports}       & integer       & Number of exported functions                                                                                           \\
		\texttt{imports}            & list          & List of imported function names (JSON array)                                                                           \\
		\texttt{exports}            & list          & List of exported function names (JSON array)                                                                           \\
		\\
		\multicolumn{3}{l}{\textbf{Complexity Metrics}}                                                                                                                      \\
		\texttt{entropy}            & float         & Shannon entropy (0.0--8.0, higher indicates compression/encryption)                                                    \\
		\\
		\multicolumn{3}{l}{\textbf{Tokenization}}                                                                                                                            \\
		\texttt{tokens}             & list[int]     & Pre-computed BPE token IDs (variable length, max 3.1M)                                                                 \\
		\texttt{token\_count}       & integer       & Total number of tokens in sequence                                                                                     \\
		\texttt{unique\_tokens}     & integer       & Number of distinct token IDs used                                                                                      \\
		\texttt{compression\_ratio} & float         & Bytes per token (file\_size / token\_count)                                                                            \\
		\\
		\multicolumn{3}{l}{\textbf{Labels}}                                                                                                                                  \\
		\texttt{is\_malware}        & boolean       & Malware label (true for malware, false for benign)                                                                     \\
		\texttt{has\_tokens}        & boolean       & Indicates if tokenization succeeded                                                                                    \\
		\\
		\multicolumn{3}{l}{\textbf{Parser Diagnostics}}                                                                                                                      \\
		\texttt{parse\_status}      & string        & Parsing result: \texttt{success}, \texttt{partial}, or \texttt{failed}                                                 \\
		\texttt{parse\_warnings}    & list          & List of parsing warnings or errors (JSON array)                                                                        \\
		\bottomrule
	\end{tabular}
\end{table}

\section{Educator's Safety Guide}
\label{app:educator_safety}

This appendix provides high-level guidance for instructors and teaching staff who wish to incorporate \datasetname into coursework. It is not a substitute for institutional policies or legal advice, but rather a practical companion to Section~\ref{sec:access} and Section~\ref{sec:limitations}.

\subsection{Goals and Scope}

The primary goals of this guide are to:
\begin{itemize}
\item Help instructors design assignments and projects that use \datasetname safely.
\item Provide starting points for conversations with institutional stakeholders (IT security, risk management, IRBs, and department leadership).
\item Outline common patterns for classroom infrastructure that reduce risk when working with malware.
\end{itemize}

The recommendations below assume that students may have heterogeneous hardware and that some work will occur on personal devices, university-managed machines, or cloud resources.

\subsection{Recommended Technical Setup (High Level)}

At a minimum, we recommend that courses using \datasetname:
\begin{itemize}
\item Rely on virtual machines, containers, or managed lab machines rather than bare-metal installs on student laptops.
\item Use network-isolated or tightly firewalled environments for any work that involves access to raw binaries.
\item Encourage use of streaming access to the tokenized dataset for ML-focused assignments, so that many students never need to download raw binaries at all.
\end{itemize}

Concrete configurations (hypervisor choice, network topology, and monitoring) should be tailored to local policies and infrastructure, and are beyond the scope of this high-level guide.

\subsection{Course Policy and Student Expectations}

Instructors should clearly communicate expectations around:
\begin{itemize}
\item \textbf{Acceptable use}: \datasetname is to be used only for coursework and defensive research; offensive activities are prohibited.
\item \textbf{Environment restrictions}: Where and how students may work with the data (e.g., department lab VMs only, no raw binaries on personal laptops).
\item \textbf{Reporting procedures}: How students should report suspected incidents (e.g., accidental execution, suspicious network activity).
\end{itemize}

Many institutions require written acceptable-use or data-handling agreements for security courses. We recommend adapting existing templates from your institution to explicitly mention the use of real malware samples.

\subsection{Institutional Coordination}

Before deploying assignments that use \datasetname, instructors should:
\begin{itemize}
\item Notify or consult with relevant IT security and risk management offices.
\item Verify whether local policy requires formal approval (e.g., from an IRB or equivalent ethics committee) for work involving malware.
\item Document the technical and procedural mitigations in place (e.g., network isolation, logging, restricted access to raw binaries).
\end{itemize}

Providing a concise summary of the dataset (including its public provenance and static-analysis focus) alongside this appendix can help non-technical reviewers understand the risk profile.

\subsection{Incident Response Considerations}

Even with precautions, mistakes can happen. Courses should have a basic incident response plan that covers:
\begin{itemize}
\item Who students should contact if they suspect malware has executed outside the intended environment.
\item What short-term steps to take (e.g., disconnecting a machine from the network, preserving logs).
\item How to coordinate with institutional IT/security teams for follow-up.
\end{itemize}

The exact procedures will vary by organization; instructors should work with local experts to define and document an appropriate plan before the course begins.

\end{document}